\documentclass[prl,aps,twocolumn,superscriptaddress,showpacs,preprintnumbers,amssymb]{revtex4-1}
\usepackage{graphicx}
\usepackage{dcolumn}
\usepackage{bm}
\usepackage{xcolor}
\usepackage{amsmath} 

\usepackage{hyperref}
\hypersetup{colorlinks=true, anchorcolor=blue, citecolor=blue, linkcolor=blue, filecolor=blue, urlcolor=blue, pdfpagemode=FullScreen}

\begin{document}

\title{Soft-phonon and charge-density-wave formation in nematic BaNi$_2$As$_2$}

\author{S. M. Souliou}
\email{michaela.souliou@kit.edu}
\affiliation{Institute for Quantum Materials and Technologies, Karlsruhe Institute of Technology, D-76021 Karlsruhe}
\author{T. Lacmann}
\affiliation{Institute for Quantum Materials and Technologies, Karlsruhe Institute of Technology, D-76021 Karlsruhe}
\author{R. Heid}
\affiliation{Institute for Quantum Materials and Technologies, Karlsruhe Institute of Technology, D-76021 Karlsruhe}	
\author{C. Meingast}
\affiliation{Institute for Quantum Materials and Technologies, Karlsruhe Institute of Technology, D-76021 Karlsruhe}
\author{M. Frachet}
\affiliation{Institute for Quantum Materials and Technologies, Karlsruhe Institute of Technology, D-76021 Karlsruhe}
\author{L. Paolasini}
\affiliation{ESRF – The European Synchrotron, 71, avenue des Martyrs, CS 40220 F-38043 Grenoble Cedex 9}
\author{A.-A. Haghighirad}
\affiliation{Institute for Quantum Materials and Technologies, Karlsruhe Institute of Technology, D-76021 Karlsruhe}
\author{M. Merz}
\affiliation{Institute for Quantum Materials and Technologies, Karlsruhe Institute of Technology, D-76021 Karlsruhe}
\affiliation{Karlsruhe Nano Micro Facility (KNMFi), Karlsruhe Institute of Technology, 76344 Eggenstein-Leopoldshafen\\}
\author{A. Bosak}
\affiliation{ESRF – The European Synchrotron, 71, avenue des Martyrs, CS 40220 F-38043 Grenoble Cedex 9}
\author{M. Le Tacon}
\email{matthieu.letacon@kit.edu}	
\affiliation{Institute for Quantum Materials and Technologies, Karlsruhe Institute of Technology, D-76021 Karlsruhe}

\date{\today}

\begin{abstract}
We use diffuse and inelastic x-ray scattering to study the formation of an incommensurate charge-density-wave (I-CDW) in BaNi$_2$As$_2$, a candidate system for charge-driven electronic nematicity. Intense diffuse scattering is observed around the modulation vector of the I-CDW, $Q_{I-CDW}$. It is already visible at room temperature and collapses into superstructure reflections in the long-range ordered state where a small  
orthorhombic distortion occurs.
A clear dip in the dispersion of a low-energy transverse optical phonon mode is observed around  $Q_{I-CDW}$. 
The phonon continuously softens upon cooling, ultimately driving the transition to the I-CDW state. The transverse character of the soft-phonon branch elucidates the complex pattern of the I-CDW satellites observed in the current and earlier studies  and settles the debated unidirectional nature of the I-CDW. The phonon instability and its reciprocal space position is well captured by our ${ab}$ ${initio}$ calculations. These however indicate that neither Fermi surface nesting, nor enhanced momentum-dependent electron-phonon coupling can account for the I-CDW formation, demonstrating its unconventional nature.
\end{abstract}

\maketitle
Electronic nematicity can emerge out of the fluctuations of an electronic phase characterized by a multi-component order parameter, such as those encountered in e.g. unconventional superconductors or density-waves materials~\cite{Fernandes2019}. This notion has been central in uncovering the physical properties of iron-based superconductors, in which the metallic parent compounds exhibit a doubly degenerate spin-density-wave (SDW), hosted on a square lattice. Electronic nematicity can in principle also emerge close to charge-density-wave (CDW) states, which are also routinely observed close to superconducting phases as in e.g. the high temperature superconducting cuprates~\cite{Achkar2016}, the transition-metal dichalcogenides~\cite{Cho2020} or the recently discovered Kagome superconductors~\cite{Neupert2022}. The nature and impact of the interplay between superconductivity, nematicity and CDW in all these materials is currently strongly debated.

This debate has recently been fueled with the observation of a six-fold enhancement of the superconducting critical temperature ($T_c$) through nematic fluctuations in a Ni-based family of compounds isostructural to the parent compound of the iron-based superconductors BaFe$_2$As$_2$~\cite{Eckberg2020, Lederer2020}. In contrast to its Fe-based cousin, no magnetic ordering has been reported in BaNi$_2$As$_2$, in which a series of structural and CDW instabilities is observed~\cite{Lee2019,Lee2021,Merz2021}. 
Moreover, while in non-superconducting BaFe$_2$As$_2$ electronic nematicity induces a tetragonal-to-orthorhombic transition at $\sim$ 137 K~\cite{Chu2010}, BaNi$_2$As$_2$ undergoes a first-order phase transition into a triclinic structure (SG P$\bar{1}$) at $T_{tri}^{cool}$ = 135 K (upon cooling)~\cite{Ronning2008,Sefat2009,Kudo2012}. Superconductivity appears below $T_c$ = 0.7 K, already without chemical doping or substitutions, and thermodynamic measurements indicate a fully gapped superconducting state~\cite{Kurita2009}.
An incommensurate CDW (I-CDW) at the ordering wavevectors $q_{I-CDW}$ = (0.28 0 0) and (0 0.28 0) has been reported by x-ray diffraction (XRD)~\cite{Lee2019,Lee2021,Merz2021} (throughout this paper, the momentum transfers are quoted in reciprocal lattice units (r.l.u.) of the tetragonal crystal structure ~\cite{SOM}). 
The corresponding superstructure reflections were observed below $\sim$ 155 K $> T_{tri}$. Below $T_{tri}$, the I-CDW is replaced by a commensurate modulation (C-CDW) with an ordering vector $q_{C-CDW}$ = (1/3 0 1/3)~\cite{Lee2019,Lee2021,Merz2021}. Although the relation between the I-CDW and the C-CDW are currently debated, a recent time-resolved optical spectroscopy study suggests that the C-CDW evolves from the I-CDW by gaining additional periodicity along the $c$-axis~\cite{Grigorev2021}. 
 The I-CDW therefore appears to be the primary instability of the high-temperature tetragonal phase of BaNi$_2$As$_2$, and could play a role similar to that of magnetism in the iron-based superconductors and in particular yield a form of charge-order-induced electronic nematicity.

\begin{figure*}[!ht]
\centering
\includegraphics[width=\textwidth]{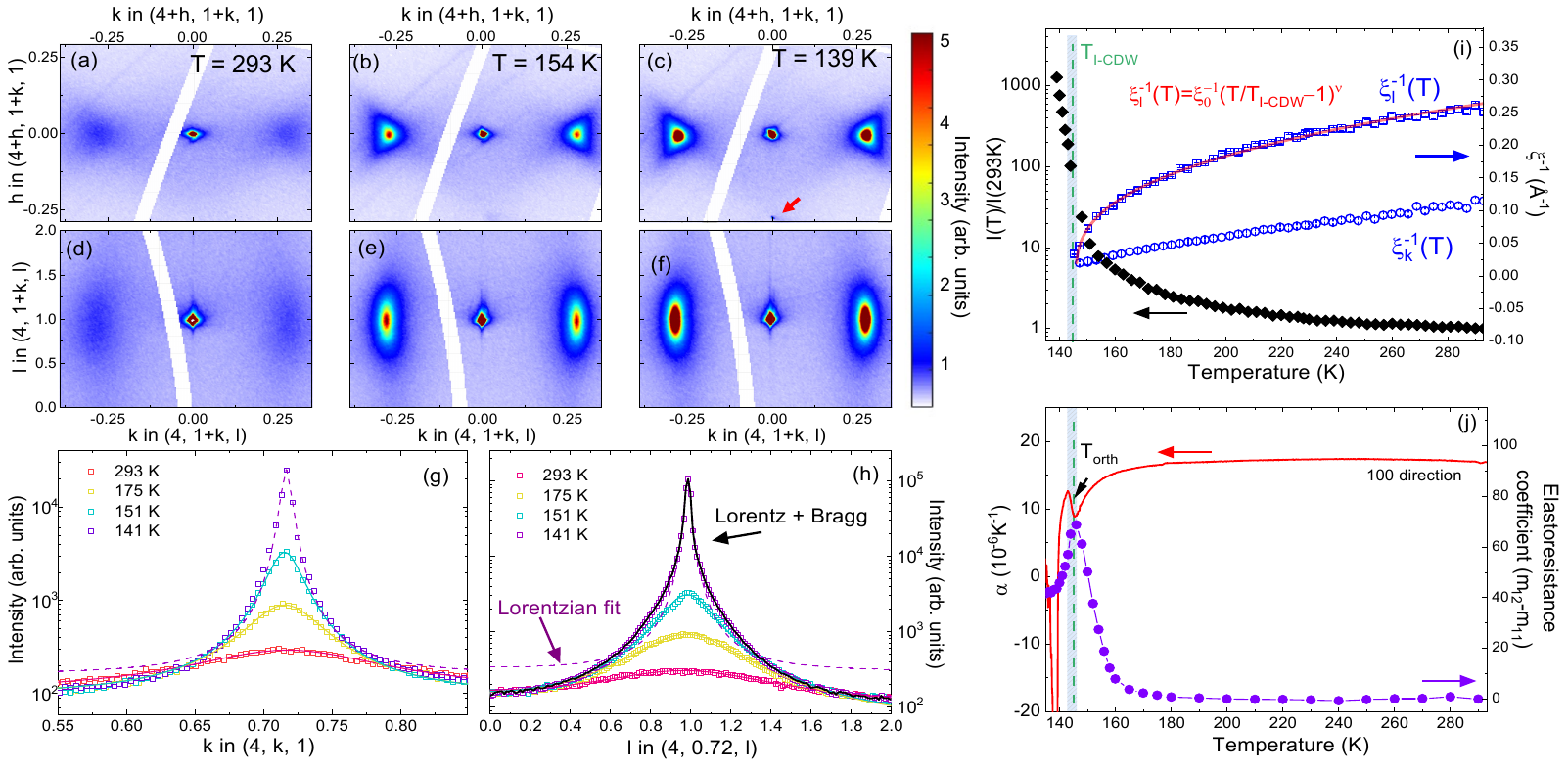}
\caption{Intensity maps of the HK1 reciprocal lattice plane around the $\Gamma_{411}$ Brillouin zone center at (a) 293 K, (b) 154 K, and (c) 139 K. Intensity maps of the 4KL plane at (d) 293 K, (e) 154 K, and (f) 139 K. (g) Cut of the DS maps across $Q_{I-CDW}$ at 293, 175, 151 and 141 K along the [0k0] direction. Lines are fit to the data using a Lorentzian model that reproduces the lineshape accurately down to $T_{I-CDW}$. 
(h) Same as (g) but for cuts along the [00l] direction. (i) T-dependence of the normalized intensity of the I-CDW superstructure peak and of the inverse I-CDW correlation length in the in-plane [0k0] and out-of-plane [00l] directions. (j) Thermal expansion coefficient along the a-axis, $\alpha$=1/L·dL/dT (L stands for the lenght of the sample) and B$_{1g}$ symmetry-resolved $m_{12}-m_{11}$ elastoresistance coefficient from ref.~\cite{Meingast2022,Frachet2022}.} 
\label{Fig1}
\end{figure*}
Along these lines, a large B$_{1g}$ nematic susceptibility and a strain-hysteretic behavior in the presence of the I-CDW order have been reported in a recent series of elastoresistivity measurements on BaNi$_2$As$_2$~\cite{Eckberg2020, Frachet2022}. 
These observations have been interpreted as tetragonal symmetry breaking in the B$_{1g}$ symmetry channel above $T_{tri}$ and suggested a link between the I-CDW and nematicity. 
Additionally, high-resolution dilatometry revealed that at $T_{orth}$ = 142 K (i.e. below the reported appearance temperature of the I-CDW satellites but above $T_{tri}$), a small orthorhombic distortion of the lattice occurs~\cite{Merz2021,Meingast2022}.
Moreover, based on the absence of a precursor response in the electronic nematic susceptibility, a lattice-driven (rather than electronic) origin of the nematicity in BaNi$_2$As$_2$ was suggested~\cite{Eckberg2020}. This conclusion was later challenged by the unsettled issue of the I-CDW unidirectionality~\cite{Lee2021}. 

These observations emphasize a tight connection between I-CDW, nematicity and structural transitions in BaNi$_2$As$_2$. As these phenomena have recently proven ubiquitous in many quantum materials~\cite{Fernandes2019,Achkar2016,Cho2020,Neupert2022}, gaining understanding of the mechanism underpinning their formation is of central importance.
Even though an earlier theoretical calculation has not revealed lattice instabilities in this system~\cite{Subedi2008}, the study of the lattice dynamics through the dispersion of phonons is most relevant. In the case of BaNi$_2$As$_2$, however, it has only been limited to zone center Raman active phonons~\cite{Yao2022} and no information is available on the phononic behavior around $q_{I-CDW}$. 
Here, we address this issue by studying the low-energy lattice dynamics of BaNi$_2$As$_2$, using a combination of diffuse and inelastic x-ray scattering with \textit{ab initio} calculations. 
We observe intense temperature-dependent CDW-related diffuse scattering signal and a pronounced anomaly of a transverse phonon branch predicted to be unstable by our density-functional perturbation theory (DFPT) calculations, in sharp contrast to \textit{e.g.} the case of cuprates~\cite{Kim2018}. This anomaly deepens upon cooling towards the I-CDW long-range ordering temperature, unambiguously establishing a soft-phonon driven condensation. The transverse character of the mode further provides a natural explanation for the XRD pattern of the I-CDW~\cite{Lee2021, Merz2021}. In agreement with recent studies of the Fermi surface (FS)~\cite{Guo2022, Pavlov2021}, nesting-based  mechanisms appear irrelevant for BaNi$_2$As$_2$, and possible alternative microscopic mechanisms are discussed.

We used high-quality BaNi$_2$As$_2$ single crystals grown by a self-flux method and characterized by XRD~\cite{SOM}. The diffuse scattering (DS) and inelastic x-ray scattering (IXS) experiments were performed at the ID28 beamline of the European Synchrotron Radiation Facility (ESRF). The IXS measurements were conducted with $\sim$3 meV energy resolution~\cite{Girard2019,Krisch2006}. The DS data were recorded with a Pilatus3 X 1M detector in shutterless mode. 
More information on the samples, and the experimental and computational methods can be found in the Supplemental Material (SM)~\cite{SOM}.

We started our investigation with a DS survey of the reciprocal space of BaNi$_2$As$_2$. DS signal around $q_{I-CDW}$ was observed next to the main Bragg reflections already at room temperature (RT).
For the detailed T-dependence, we focused on the part of the reciprocal space close to the (4 1 1) Bragg reflection (hereafter $\Gamma_{411}$), where the DS signal at $q_{I-CDW}$ is intense and, where, as we shall see below, the structure factor of an unstable phonon is strong. The DS datasets were collected upon heating from 139 K $> T_{tri}$ up to RT. Reconstructed intensity maps of the (H K 1) plane (resp. (4 K L) plane) around $\Gamma_{411}$ are presented for selected temperatures in Figs.\ref{Fig1}(a), \ref{Fig1}(b), \ref{Fig1}(c) (resp. Figs.\ref{Fig1}(d), \ref{Fig1}(e), \ref{Fig1}(f)). Further DS data are presented in the SM. 
Interestingly, throughout the temperature series and down to $\sim$ 150 K, strong DS signal is observed at $Q_{I-CDW}$=(4 1$\pm$0.28 1), but not at (4$\pm$0.28 1 1), indicative of the unidirectional and transverse character of the modulation.
At RT, cuts across $\Gamma_{411}$ - $q_{I-CDW}$ = (4 0.72 1) and along the [0k0] and [00l] directions  can be fitted with Lorentzian profiles (Figs.\ref{Fig1}(g) and \ref{Fig1}(h)) of HWHM of $\sim$ 0.08 r.l.u. and 0.47 r.l.u., respectively, corresponding to very short in-plane $\xi_{k} \sim$ 10 \AA~  and out-of-plane  $\xi_{l} \sim$ 4 \AA~ correlation lengths. 
The peak sharpens and becomes more intense at low temperatures, and cannot be fitted with a simple Lorentzian profile below $\sim$ 146 K, where the peak intensity diverges. Below this temperature, the CDW satellites are as intense as the nearby Bragg reflection. 
In Fig.\ref{Fig1}(i), we show the T-dependence of the inverse of the correlation lengths both in- and out-of-plane. Interestingly, the increase of the  $\xi_{k}$ is rather gradual, while the out-of-plane correlation length $\xi_{l}$ displays a critical divergence towards $T_{I-CDW}$ - the onset temperature of the long-range order, as defined below. This relates to the recently reported shrinking of the c/a ratio upon cooling in this system~\cite{Meingast2022}. A fitting of the T-dependence  $\xi_{l}(T)= \xi_{0}(T/T_{I-CDW}-1)^{-\nu}$ yields a value of $\nu = 0.36 \pm 0.01$ and $T_{I-CDW} = 146 \pm 1$K (dashed line in Figs.~\ref{Fig1}(i) and (j)). This indicates the formation of a true long-range order at $T_{I-CDW}$. This temperature also corresponds (Fig.~\ref{Fig1}(j)) to a maximum of the B$_{1g}$ elasto-resistance coefficient~\cite{Frachet2022} and to a jump of the thermal expansion coefficient $\alpha$ (corresponding to the shaded area centered around 145 K in Figs.~\ref{Fig1}(i) and (j)), typical of a second order phase transition, hereby confirming that the small orthorhombic distortion at $T_{orth}$ is a by-product of the formation of the long-range I-CDW~\cite{Meingast2022, Merz2021}. Finally, below $T_{I-CDW}$, we note the appearance of a much weaker and sharp feature at (4.28 1 1), i.e. in the longitudinal direction (red arrow in Fig. \ref{Fig1}(c)).

\begin{figure}[b]
\centering
\includegraphics[width=0.5\textwidth]{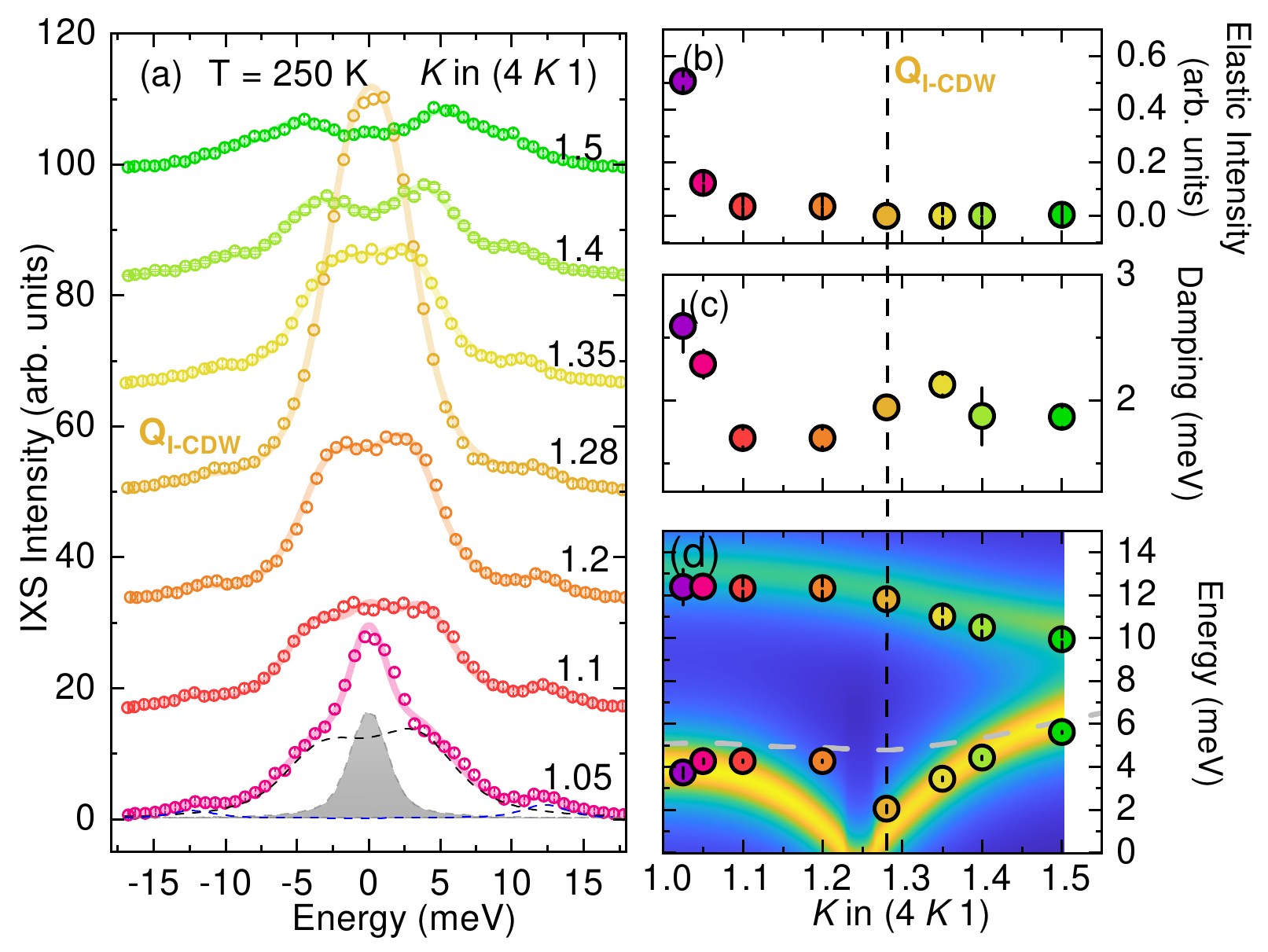}
\caption{(a) IXS spectra taken from $\Gamma_{411}$ along the [0k0] direction at 250 K (vertically shifted for clarity). Thick solid lines correspond to fit of the data (SM). Fitting details are shown for the spectra at Q = (4 1.05 1). The dashed lines represent the two individual phonons (both the Stokes and anti-Stokes parts are included). The gray line represents the elastic contribution. Momentum dependence (b) of the elastic line intensity, (c) of the linewidth of the low energy phonon and (d) of the energy of the two phonons shown in (a) at 250 K. The colormap represents the scattering intensity along [0k0] in the zone centered around $\Gamma_{411}$ calculated by DFPT. The gray dashed line is the phonon dispersion calculated using a large gaussian smearing that simulates high temperatures and stabilizes the tetragonal phase~\cite{SOM}}.
\label{Fig2}
\end{figure}

\begin{figure}[t]
\centering
\includegraphics[width=0.5\textwidth]{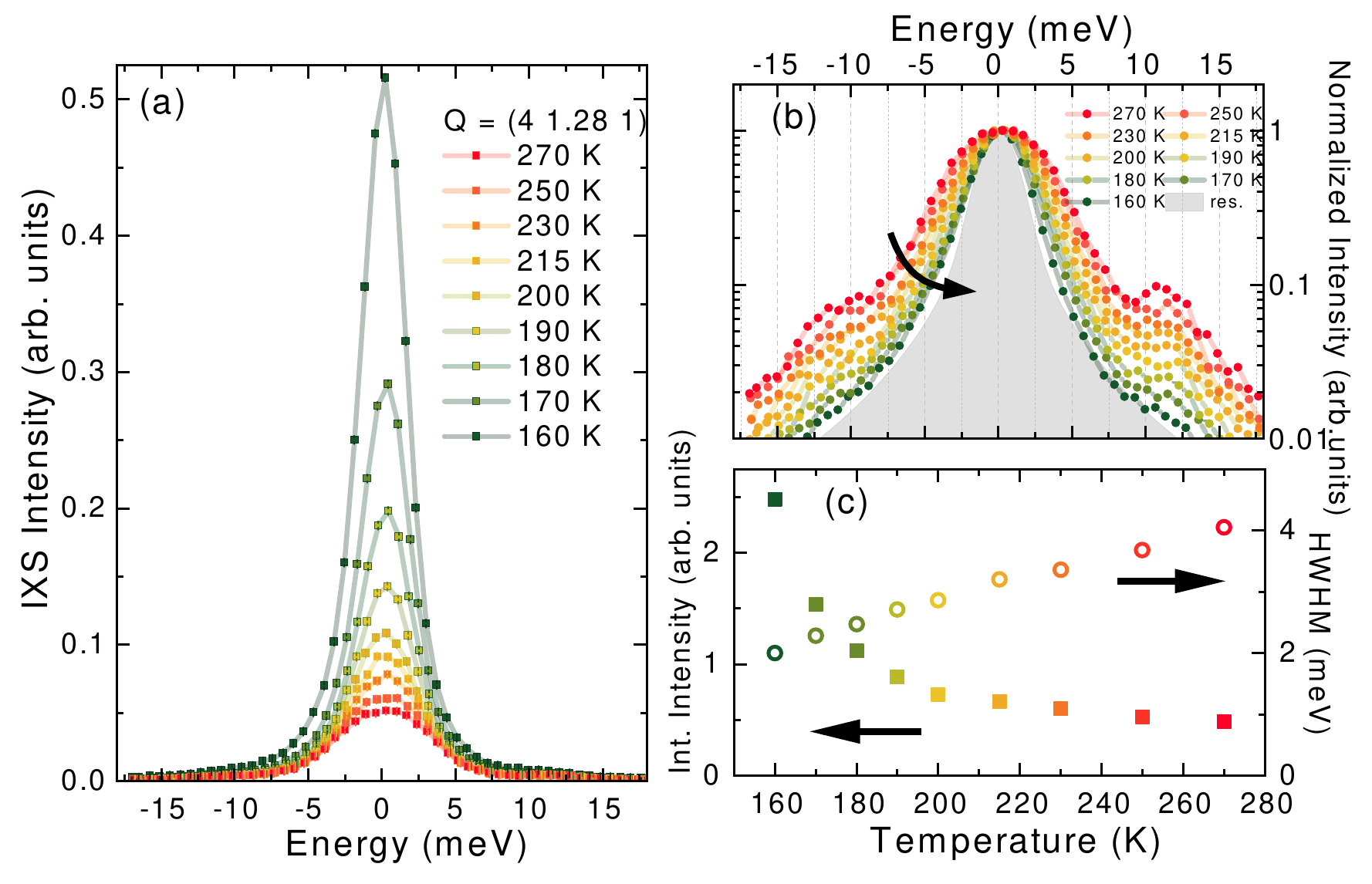}
\caption{(a) T-dependence of the IXS spectra at $Q_{I-CDW}$. (b) Normalized IXS intensity at $Q_{I-CDW}$ as a function of temperature. (c) T-dependence of the integrated spectral weight and of the HWHM of the quasielastic line at $Q_{I-CDW}$.} 
\label{Fig3}
\end{figure}

DS signal can in principle arise from static disorder (e.g. pinned CDW domains) or from soft-phonons.
In order to gain more insights on the origin of the observed signal, we have carried out a series of energy-resolved IXS experiments.
In Fig. \ref{Fig2}(a), we show a selection of inelastic scans taken at 250 K for various momenta along the [0k0] direction across $Q_{I-CDW}$= $\Gamma_{411}$ + $q_{I-CDW}$ = (4 1.28 1). In the investigated energy range, we identify the elastic line centered around 0 meV energy transfer and two phonons at energy transfers $\sim$ 4 and $\sim$ 12.5 meV both on the positive (Stokes) and negative (anti-Stokes) sides.
The IXS spectra are analyzed by fitting with damped harmonic oscillator (DHO) lineshapes, convoluted with the experimental resolution function~\cite{SOM}, and a resolution limited elastic line as illustrated in Fig. \ref{Fig2}(a). The number of phonons and their frequencies are in good agreement with our DFPT calculations (Fig. \ref{Fig2}(d) and SM~\cite{SOM}). The two modes are optical phonons dispersing from doubly degenerate, in-plane polarized, E$_g$ Raman- and E$_u$ infrared-active zone center modes, respectively. The higher energy phonon disperses continuously towards lower energies between the zone center and the zone edge, while the lower energy phonon softens around $Q_{I-CDW}$ (Fig. \ref{Fig2}(d)). This anomaly is also captured by our calculations, which predict an instability of the lower optical phonon branch at a wavevector very close to $Q_{I-CDW}$. Most importantly in the chosen Brillouin zone, the calculated scattering intensity of the unstable phonon is strong, ensuring that the IXS experiment is carried out in the best condition to observe the mode anomaly if it really exists. This instability can be suppressed by simulating qualitatively the effect of a very high temperature on the phonon dispersion using a large gaussian smearing of the Fermi distribution function in the \textit{ab initio} calculation~\cite{SOM}. The resulting dispersion of the low energy optical mode is shown in white in panel (d) of Fig. \ref{Fig2}. As seen in Fig. \ref{Fig2}(a), the spectrum at  $Q_{I-CDW}$ essentially consists of a quasi-elastic line, which is however much broader than the experimental energy resolution (see also Fig. \ref{Fig3}). To analyze the IXS spectra, we used a fitting procedure in which the energy-integrated spectral-weight of the DHO lines describing the phonons was constrained. It is based on the fact that this spectral-weight is proportional to the ratio of the phonon structure factor (which is essentially constant over the investigated range of momenta) to the square of the oscillator´s frequency (see the SM for more details). This revealed that away from the Brillouin zone center (and in particular around $Q_{I-CDW}$), the spectra at 250 K can essentially be described using only phonons, with a negligible elastic scattering contribution (Fig. \ref{Fig2}(b)). We note a significant broadening of the phonon lineshape (Fig. \ref{Fig2}(c)) around $Q_{I-CDW}$, where the phonon is also particularly soft and close to being overdamped. It is worth mentioning here that the phonon on the entire branch is strongly damped (HWHM $\sim$ 2 meV). This agrees with Raman data showing that the E$_g$ mode out-of-which the branch emerges is particularly broad~\cite{Yao2022} already at RT. Within our resolution we did not observe any anomaly in the dispersion of the higher energy phonon. 

In Fig. \ref{Fig3}(a) we show the T-dependence of the IXS spectra at $Q_{I-CDW}$. The broad quasielastic line observed at high temperatures strongly grows and narrows upon cooling. This can be better seen in Fig. \ref{Fig3}(b), where we plot on a logarithmic intensity scale the normalized intensities recorded between 270 and 160 K. The T-dependencies of the integrated intensity of the spectra as well as the HWHM of the quasi-elastic line are reported in Fig.\ref{Fig3}(c). 
At 160 K, the spectrum remains slightly broader than the instrumental resolution (Fig. \ref{Fig3}(b)). This observation unambiguously evidences a continuous softening of the lowest energy optical phonon at $Q_{I-CDW}$ upon cooling. As discussed in the SM~\cite{SOM}, however, the combination of a very soft energy with a broad linewidth and an intense elastic line did not allow us to extract with sufficient confidence the T-dependence of the soft-mode energy. We note however that down to at least 170K, a model with a finite phonon frequency always yields a better fit to the data than a model in which the phonon is completely soft and overdamped~\cite{Song2022}.

The IXS data associate the temperature-dependent DS signal with the softening of a low energy optical phonon at $Q_{I-CDW}$, and the strong increase of the DS intensity below  $T_{I-CDW}$ with its complete condensation. As discussed earlier, above $T_{I-CDW}$ DS is only observed  along the $K$ direction, at $Q_{I-CDW}$=(4 1$\pm$0.28 1). In agreement, our IXS measurements along the orthogonal direction (4+h 1 1) did not reveal any phonon anomaly at (4$\pm$0.28 1 1)- see also the SM~\cite{SOM}. The IXS data along the two tetragonal directions in combination with the results of our DFPT calculations demonstrate that the soft-phonon belongs to a transverse optical phonon branch, which explains the  transverse character of the modulation seen in DS and accounts for the 'square' pattern formed by the I-CDW reflections~\cite{Merz2021}. The satellites are consistently the ones in the direction in which the transverse phonon branch is probed, and are absent in the longitudinal directions.
The distinction between longitudinal and transverse geometries is less clear in reciprocal lattice planes with higher L indexes, where they mix, as in the L=5 case of the data presented in ref. ~\cite{Lee2021}. However, as previously mentioned, weak satellites along the longitudinal (4+h 1 1) direction are seen at low temperatures. Unlike the transverse satellites, these are strictly momentum-resolution limited, are not visible above $T_{I-CDW}$ nor associated with a soft-phonon~\cite{SOM} and therefore rather relate to the long-range CDW order.

\begin{figure}[b]
\centering
\includegraphics[width=0.48\textwidth]{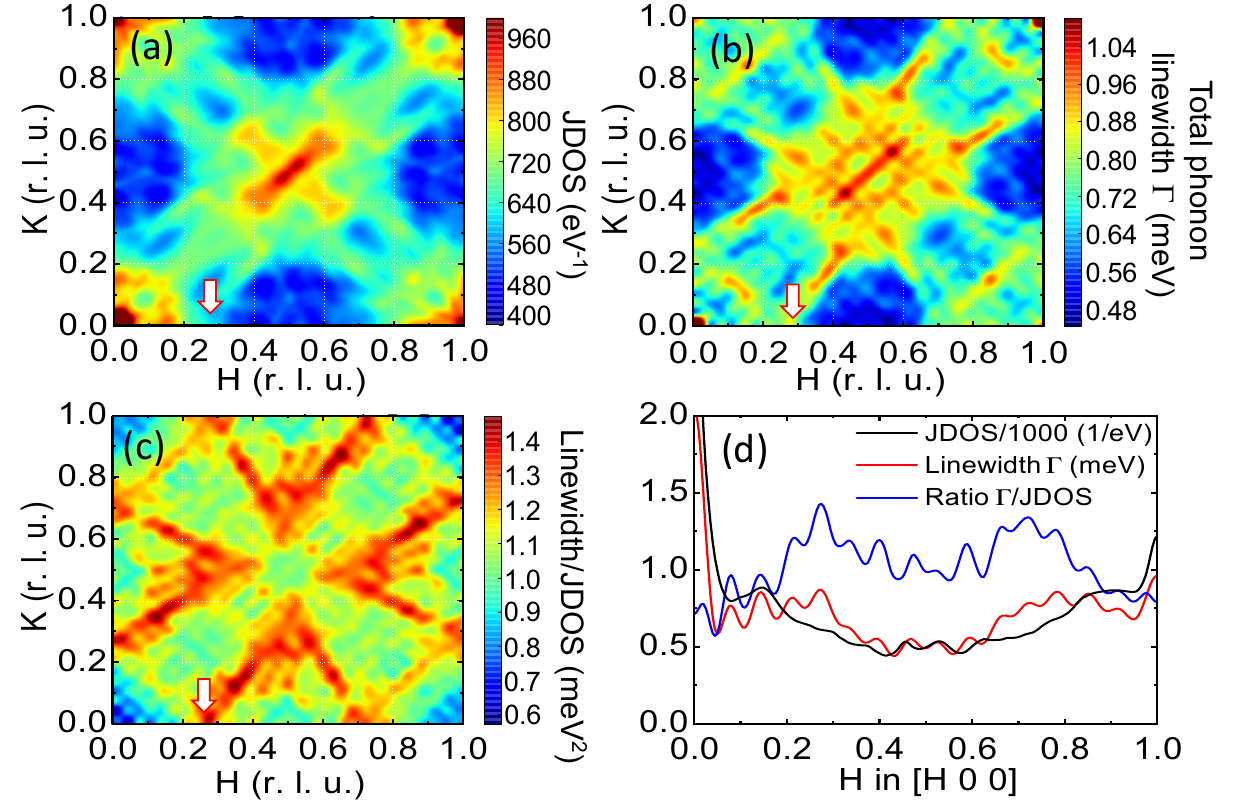}
\caption{Calculated momentum-dependent electronic and electron-phonon coupling (EPC) quantities at the 2D plane $k_z=0$ (see SM for their definitions): (a) Joint density of states (JDOS); (b) Total linewidth, i.e., phonon linewidth summed over all phonon branches; (c) Ratio of total linewidth and JDOS, which represents an average EPC strength. The arrow in (a)-(c) indicates $q_{I-CDW}$ (d) Line cuts along the [100] direction for the quantities shown in (a)-(c).}
\label{Fig4}
\end{figure}	

Our experimental results unambiguously establish that the I-CDW formation is driven by the softening of a transverse phonon, yet the origin of this phenomenon remains unclear. Recent time-resolved optical spectroscopy data have shown the resilience of the CDW order against optical excitation up to very high fluences, suggesting an unconventional nature~\cite{Grigorev2021}.
This is in line with previous electronic structure calculations and angle resolved photoemission spectroscopy (ARPES) experiments, which have not identified nesting features in the FS of BaNi$_2$As$_2$ above $T_{Tri}$~\cite{Subedi2008,Zhou2011,Guo2022, Pavlov2021}, ruling out weak-coupling mechanisms~\cite{Peierls2001,Gruener2018}.
This is confirmed by the absence of nesting features around $q_{I-CDW}$ in our calculated joint density-of-states (JDOS) (Fig.~\ref{Fig4}(a)). We have evaluated further the $q$-dependence of the electron-phonon coupling (EPC) matrix elements by calculating the ratio of the total (summed over all branches) phonon linewidth (Fig.~\ref{Fig4}(b)) over the JDOS (Fig.~\ref{Fig4}(c)). If a shallow maximum - best seen in the cuts along the [100] direction presented in Fig.~\ref{Fig4}(d) - appears around $q_{I-CDW}$ and suggests a local enhancement of the EPC matrix elements, it is neither pronounced nor sharp in momentum space. Furthermore the soft-phonon does not even appear as one of the main contributors to the EPC~\cite{SOM}. Consequently, alternative CDW formation mechanisms based on momentum-dependent EPC akin to what has been reported in e.g.  dichalcogenides~\cite{Weber2011,Weber2011a} or more recently in LaAgSb$_2$, in combination with nesting of a subset of Fermi sheets~\cite{Bosak2021}, can confidently be ruled out as well for BaNi$_2$As$_2$. 

As previously discussed, the unstable phonon both in the calculation and in the experiment is transversely polarized, an aspect which is not captured by the above analysis. Furthermore the soft branch disperses from the E$_g$ mode which exhibits an anomalously large splitting~\cite{Yao2022} onsetting above $T_{I-CDW}$. This effect can be accounted for by a strong coupling between this mode and an Ising nematic degree of freedom - 
possibly of orbital nature (given the absence of magnetism) - with the B$_{1g}$ symmetry. Additional insights on the nature of the CDW order come from recent ARPES experiments, which found evidence for orbital-dependent band renormalization at low temperatures related to anisotropic Ni-Ni bond ordering~\cite{Noda2017}. Along the same lines, the formation of Ni-Ni dimers was recently inferred from a refinement of low temperature XRD data, supported by near-edge x-ray absorption fine structure experiments showing that charge fluctuations between out-of-plane and in-plane orbitals are present already above $T_{I-CDW}$~\cite{Merz2021}. This suggests that mechanisms involving orbital degrees of freedom, encompassing orbital-dependent EPC~\cite{Flicker_NatCom2015} or orbitally driven Peierls states~\cite{Khomskii2005,Streltsov2014,Khomskii2020} might be at play here. A determination of the orbital texture of the I-CDW order by means of complementary techniques, such as resonant x-ray scattering, would be required in order to clarify the role of orbital fluctuations in the stabilization of the I-CDW. 

In summary, our study revealed intense DS and a pronounced anomaly of a transverse optical phonon at the I-CDW ordering vector of BaNi$_2$As$_2$. 
As previously conjectured~\cite{Merz2021, Meingast2022}, our detailed comparison with thermodynamic data confirm that the small orthorhombic distortion is a direct consequence of the formation of the long-range uniaxial I-CDW state, which is itself driven by the softening of this transverse phonon. Our calculations show that the instability of the mode can neither be associated with FS nesting nor with local enhancement of the EPC, and supports an unconventional mechanism. We conclude by noting that thermodynamic measurements indicated that the five-fold increase of the superconducting $T_c$ at Phosphorus doping higher than 7\% is related to a giant phonon softening occurring when the triclinic transition is completely suppressed~\cite{Kudo2012, Meingast2022}. Future DS and IXS measurements in this doping regime will be valuable in elucidating the relevance of nematicity, the structural transition and the CDW ordering in the superconducting pairing. More generally, the approach used here, combining first principle calculations, DS and IXS, is most relevant to address - or revisit - the formation mechanism of recently discovered long-range I-CDW in systems such as overdoped cuprates~\cite{Peng2020,TamNatCom2022} or Kagome superconductors~\cite{LiPRX2021, Neupert2022}.

\textit{Note added.} During the completion of this manuscript, an IXS study of a phonon softening associated with the I-CDW in BaNi$_2$As$_2$ has been reported ~\cite{Song2022}.

\section*{Acknowledgements}
This work was funded by the Deutsche Forschungsgemeinschaft (DFG, German Research Foundation) - TRR 288 - 422213477 (project B03). S.M.S. acknowledges funding by the DFG – Projektnummer 441231589.
R.H. acknowledges support by the state of Baden-W\"{u}rttemberg through bwHPC. We thank P. Dai, T. Forrest, J. Paglione, J. Schmalian, F. Weber K. Willa and M. Yi for fruitful discussions.

\bibliographystyle{apsrev4-1}

\begin{thebibliography}{51}%
\makeatletter
\providecommand \@ifxundefined [1]{%
 \@ifx{#1\undefined}
}%
\providecommand \@ifnum [1]{%
 \ifnum #1\expandafter \@firstoftwo
 \else \expandafter \@secondoftwo
 \fi
}%
\providecommand \@ifx [1]{%
 \ifx #1\expandafter \@firstoftwo
 \else \expandafter \@secondoftwo
 \fi
}%
\providecommand \natexlab [1]{#1}%
\providecommand \enquote  [1]{``#1''}%
\providecommand \bibnamefont  [1]{#1}%
\providecommand \bibfnamefont [1]{#1}%
\providecommand \citenamefont [1]{#1}%
\providecommand \href@noop [0]{\@secondoftwo}%
\providecommand \href [0]{\begingroup \@sanitize@url \@href}%
\providecommand \@href[1]{\@@startlink{#1}\@@href}%
\providecommand \@@href[1]{\endgroup#1\@@endlink}%
\providecommand \@sanitize@url [0]{\catcode `\\12\catcode `\$12\catcode
  `\&12\catcode `\#12\catcode `\^12\catcode `\_12\catcode `\%12\relax}%
\providecommand \@@startlink[1]{}%
\providecommand \@@endlink[0]{}%
\providecommand \url  [0]{\begingroup\@sanitize@url \@url }%
\providecommand \@url [1]{\endgroup\@href {#1}{\urlprefix }}%
\providecommand \urlprefix  [0]{URL }%
\providecommand \Eprint [0]{\href }%
\providecommand \doibase [0]{http://dx.doi.org/}%
\providecommand \selectlanguage [0]{\@gobble}%
\providecommand \bibinfo  [0]{\@secondoftwo}%
\providecommand \bibfield  [0]{\@secondoftwo}%
\providecommand \translation [1]{[#1]}%
\providecommand \BibitemOpen [0]{}%
\providecommand \bibitemStop [0]{}%
\providecommand \bibitemNoStop [0]{.\EOS\space}%
\providecommand \EOS [0]{\spacefactor3000\relax}%
\providecommand \BibitemShut  [1]{\csname bibitem#1\endcsname}%
\let\auto@bib@innerbib\@empty
\bibitem [{\citenamefont {Fernandes}\ \emph {et~al.}(2019)\citenamefont
  {Fernandes}, \citenamefont {Orth},\ and\ \citenamefont
  {Schmalian}}]{Fernandes2019}%
  \BibitemOpen
  \bibfield  {author} {\bibinfo {author} {\bibfnamefont {R.~M.}\ \bibnamefont
  {Fernandes}}, \bibinfo {author} {\bibfnamefont {P.~P.}\ \bibnamefont {Orth}},
  \ and\ \bibinfo {author} {\bibfnamefont {J.}~\bibnamefont {Schmalian}},\
  }\href {\doibase 10.1146/annurev-conmatphys-031218-013200} {\bibfield
  {journal} {\bibinfo  {journal} {Annual Review of Condensed Matter Physics}\
  }\textbf {\bibinfo {volume} {10}},\ \bibinfo {pages} {133} (\bibinfo {year}
  {2019})}\BibitemShut {NoStop}%
\bibitem [{\citenamefont {Achkar}\ \emph {et~al.}(2016)\citenamefont {Achkar},
  \citenamefont {Zwiebler}, \citenamefont {McMahon}, \citenamefont {He},
  \citenamefont {Sutarto}, \citenamefont {Djianto}, \citenamefont {Hao},
  \citenamefont {Gingras}, \citenamefont {H\"{u}cker}, \citenamefont {Gu},
  \citenamefont {Revcolevschi}, \citenamefont {Zhang}, \citenamefont {Kim},
  \citenamefont {Geck},\ and\ \citenamefont {Hawthorn}}]{Achkar2016}%
  \BibitemOpen
  \bibfield  {author} {\bibinfo {author} {\bibfnamefont {A.~J.}\ \bibnamefont
  {Achkar}}, \bibinfo {author} {\bibfnamefont {M.}~\bibnamefont {Zwiebler}},
  \bibinfo {author} {\bibfnamefont {C.}~\bibnamefont {McMahon}}, \bibinfo
  {author} {\bibfnamefont {F.}~\bibnamefont {He}}, \bibinfo {author}
  {\bibfnamefont {R.}~\bibnamefont {Sutarto}}, \bibinfo {author} {\bibfnamefont
  {I.}~\bibnamefont {Djianto}}, \bibinfo {author} {\bibfnamefont
  {Z.}~\bibnamefont {Hao}}, \bibinfo {author} {\bibfnamefont {M.~J.~P.}\
  \bibnamefont {Gingras}}, \bibinfo {author} {\bibfnamefont {M.}~\bibnamefont
  {H\"{u}cker}}, \bibinfo {author} {\bibfnamefont {G.~D.}\ \bibnamefont {Gu}},
  \bibinfo {author} {\bibfnamefont {A.}~\bibnamefont {Revcolevschi}}, \bibinfo
  {author} {\bibfnamefont {H.}~\bibnamefont {Zhang}}, \bibinfo {author}
  {\bibfnamefont {Y.-J.}\ \bibnamefont {Kim}}, \bibinfo {author} {\bibfnamefont
  {J.}~\bibnamefont {Geck}}, \ and\ \bibinfo {author} {\bibfnamefont {D.~G.}\
  \bibnamefont {Hawthorn}},\ }\href {\doibase 10.1126/science.aad1824}
  {\bibfield  {journal} {\bibinfo  {journal} {Science}\ }\textbf {\bibinfo
  {volume} {351}},\ \bibinfo {pages} {576} (\bibinfo {year}
  {2016})}\BibitemShut {NoStop}%
\bibitem [{\citenamefont {woo Cho}\ \emph {et~al.}(2020)\citenamefont {woo
  Cho}, \citenamefont {Lyu}, \citenamefont {An}, \citenamefont {Han},
  \citenamefont {Lo}, \citenamefont {Ng}, \citenamefont {Hu}, \citenamefont
  {Gao}, \citenamefont {Li}, \citenamefont {Huang}, \citenamefont {Wang},
  \citenamefont {Schmalian},\ and\ \citenamefont {Lortz}}]{Cho2020}%
  \BibitemOpen
  \bibfield  {author} {\bibinfo {author} {\bibfnamefont {C.}~\bibnamefont {woo
  Cho}}, \bibinfo {author} {\bibfnamefont {J.}~\bibnamefont {Lyu}}, \bibinfo
  {author} {\bibfnamefont {L.}~\bibnamefont {An}}, \bibinfo {author}
  {\bibfnamefont {T.}~\bibnamefont {Han}}, \bibinfo {author} {\bibfnamefont
  {K.~T.}\ \bibnamefont {Lo}}, \bibinfo {author} {\bibfnamefont {C.~Y.}\
  \bibnamefont {Ng}}, \bibinfo {author} {\bibfnamefont {J.}~\bibnamefont {Hu}},
  \bibinfo {author} {\bibfnamefont {Y.}~\bibnamefont {Gao}}, \bibinfo {author}
  {\bibfnamefont {G.}~\bibnamefont {Li}}, \bibinfo {author} {\bibfnamefont
  {M.}~\bibnamefont {Huang}}, \bibinfo {author} {\bibfnamefont
  {N.}~\bibnamefont {Wang}}, \bibinfo {author} {\bibfnamefont {J.}~\bibnamefont
  {Schmalian}}, \ and\ \bibinfo {author} {\bibfnamefont {R.}~\bibnamefont
  {Lortz}},\ }\href@noop {} {\  (\bibinfo {year} {2020})},\ \Eprint
  {http://arxiv.org/abs/2003.12467} {arXiv:2003.12467} \BibitemShut {NoStop}%
\bibitem [{\citenamefont {Neupert}\ \emph {et~al.}(2022)\citenamefont
  {Neupert}, \citenamefont {Denner}, \citenamefont {Yin}, \citenamefont
  {Thomale},\ and\ \citenamefont {Hasan}}]{Neupert2022}%
  \BibitemOpen
  \bibfield  {author} {\bibinfo {author} {\bibfnamefont {T.}~\bibnamefont
  {Neupert}}, \bibinfo {author} {\bibfnamefont {M.~M.}\ \bibnamefont {Denner}},
  \bibinfo {author} {\bibfnamefont {J.-X.}\ \bibnamefont {Yin}}, \bibinfo
  {author} {\bibfnamefont {R.}~\bibnamefont {Thomale}}, \ and\ \bibinfo
  {author} {\bibfnamefont {M.~Z.}\ \bibnamefont {Hasan}},\ }\href {\doibase
  10.1038/s41567-021-01404-y} {\bibfield  {journal} {\bibinfo  {journal}
  {Nature Physics}\ }\textbf {\bibinfo {volume} {18}},\ \bibinfo {pages} {137}
  (\bibinfo {year} {2022})}\BibitemShut {NoStop}%
\bibitem [{\citenamefont {Eckberg}\ \emph {et~al.}(2020)\citenamefont
  {Eckberg}, \citenamefont {Campbell}, \citenamefont {Metz}, \citenamefont
  {Collini}, \citenamefont {Hodovanets}, \citenamefont {Drye}, \citenamefont
  {Zavalij}, \citenamefont {Christensen}, \citenamefont {Fernandes},
  \citenamefont {Lee}, \citenamefont {Abbamonte}, \citenamefont {Lynn},\ and\
  \citenamefont {Paglione}}]{Eckberg2020}%
  \BibitemOpen
  \bibfield  {author} {\bibinfo {author} {\bibfnamefont {C.}~\bibnamefont
  {Eckberg}}, \bibinfo {author} {\bibfnamefont {D.~J.}\ \bibnamefont
  {Campbell}}, \bibinfo {author} {\bibfnamefont {T.}~\bibnamefont {Metz}},
  \bibinfo {author} {\bibfnamefont {J.}~\bibnamefont {Collini}}, \bibinfo
  {author} {\bibfnamefont {H.}~\bibnamefont {Hodovanets}}, \bibinfo {author}
  {\bibfnamefont {T.}~\bibnamefont {Drye}}, \bibinfo {author} {\bibfnamefont
  {P.}~\bibnamefont {Zavalij}}, \bibinfo {author} {\bibfnamefont {M.~H.}\
  \bibnamefont {Christensen}}, \bibinfo {author} {\bibfnamefont {R.~M.}\
  \bibnamefont {Fernandes}}, \bibinfo {author} {\bibfnamefont {S.}~\bibnamefont
  {Lee}}, \bibinfo {author} {\bibfnamefont {P.}~\bibnamefont {Abbamonte}},
  \bibinfo {author} {\bibfnamefont {J.~W.}\ \bibnamefont {Lynn}}, \ and\
  \bibinfo {author} {\bibfnamefont {J.}~\bibnamefont {Paglione}},\ }\href
  {\doibase 10.1038/s41567-019-0736-9} {\bibfield  {journal} {\bibinfo
  {journal} {Nature Physics}\ }\textbf {\bibinfo {volume} {16}},\ \bibinfo
  {pages} {346} (\bibinfo {year} {2020})}\BibitemShut {NoStop}%
\bibitem [{\citenamefont {Lederer}\ \emph {et~al.}(2020)\citenamefont
  {Lederer}, \citenamefont {Berg},\ and\ \citenamefont {Kim}}]{Lederer2020}%
  \BibitemOpen
  \bibfield  {author} {\bibinfo {author} {\bibfnamefont {S.}~\bibnamefont
  {Lederer}}, \bibinfo {author} {\bibfnamefont {E.}~\bibnamefont {Berg}}, \
  and\ \bibinfo {author} {\bibfnamefont {E.-A.}\ \bibnamefont {Kim}},\ }\href
  {\doibase 10.1103/physrevresearch.2.023122} {\bibfield  {journal} {\bibinfo
  {journal} {Physical Review Research}\ }\textbf {\bibinfo {volume} {2}},\
  \bibinfo {pages} {023122} (\bibinfo {year} {2020})}\BibitemShut {NoStop}%
\bibitem [{\citenamefont {Lee}\ \emph {et~al.}(2019)\citenamefont {Lee},
  \citenamefont {de~la Pe\~na}, \citenamefont {Sun}, \citenamefont {Mitrano},
  \citenamefont {Fang}, \citenamefont {Jang}, \citenamefont {Lee},
  \citenamefont {Eckberg}, \citenamefont {Campbell}, \citenamefont {Collini},
  \citenamefont {Paglione}, \citenamefont {de~Groot},\ and\ \citenamefont
  {Abbamonte}}]{Lee2019}%
  \BibitemOpen
  \bibfield  {author} {\bibinfo {author} {\bibfnamefont {S.}~\bibnamefont
  {Lee}}, \bibinfo {author} {\bibfnamefont {G.}~\bibnamefont {de~la Pe\~na}},
  \bibinfo {author} {\bibfnamefont {S.~X.-L.}\ \bibnamefont {Sun}}, \bibinfo
  {author} {\bibfnamefont {M.}~\bibnamefont {Mitrano}}, \bibinfo {author}
  {\bibfnamefont {Y.}~\bibnamefont {Fang}}, \bibinfo {author} {\bibfnamefont
  {H.}~\bibnamefont {Jang}}, \bibinfo {author} {\bibfnamefont {J.-S.}\
  \bibnamefont {Lee}}, \bibinfo {author} {\bibfnamefont {C.}~\bibnamefont
  {Eckberg}}, \bibinfo {author} {\bibfnamefont {D.}~\bibnamefont {Campbell}},
  \bibinfo {author} {\bibfnamefont {J.}~\bibnamefont {Collini}}, \bibinfo
  {author} {\bibfnamefont {J.}~\bibnamefont {Paglione}}, \bibinfo {author}
  {\bibfnamefont {F.~M.~F.}\ \bibnamefont {de~Groot}}, \ and\ \bibinfo {author}
  {\bibfnamefont {P.}~\bibnamefont {Abbamonte}},\ }\href {\doibase
  10.1103/PhysRevLett.122.147601} {\bibfield  {journal} {\bibinfo  {journal}
  {Phys. Rev. Lett.}\ }\textbf {\bibinfo {volume} {122}},\ \bibinfo {pages}
  {147601} (\bibinfo {year} {2019})}\BibitemShut {NoStop}%
\bibitem [{\citenamefont {Lee}\ \emph {et~al.}(2021)\citenamefont {Lee},
  \citenamefont {Collini}, \citenamefont {Sun}, \citenamefont {Mitrano},
  \citenamefont {Guo}, \citenamefont {Eckberg}, \citenamefont {Paglione},
  \citenamefont {Fradkin},\ and\ \citenamefont {Abbamonte}}]{Lee2021}%
  \BibitemOpen
  \bibfield  {author} {\bibinfo {author} {\bibfnamefont {S.}~\bibnamefont
  {Lee}}, \bibinfo {author} {\bibfnamefont {J.}~\bibnamefont {Collini}},
  \bibinfo {author} {\bibfnamefont {S.~X.-L.}\ \bibnamefont {Sun}}, \bibinfo
  {author} {\bibfnamefont {M.}~\bibnamefont {Mitrano}}, \bibinfo {author}
  {\bibfnamefont {X.}~\bibnamefont {Guo}}, \bibinfo {author} {\bibfnamefont
  {C.}~\bibnamefont {Eckberg}}, \bibinfo {author} {\bibfnamefont
  {J.}~\bibnamefont {Paglione}}, \bibinfo {author} {\bibfnamefont
  {E.}~\bibnamefont {Fradkin}}, \ and\ \bibinfo {author} {\bibfnamefont
  {P.}~\bibnamefont {Abbamonte}},\ }\href {\doibase
  10.1103/PhysRevLett.127.027602} {\bibfield  {journal} {\bibinfo  {journal}
  {Phys. Rev. Lett.}\ }\textbf {\bibinfo {volume} {127}},\ \bibinfo {pages}
  {027602} (\bibinfo {year} {2021})}\BibitemShut {NoStop}%
\bibitem [{\citenamefont {Merz}\ \emph {et~al.}(2021)\citenamefont {Merz},
  \citenamefont {Wang}, \citenamefont {Wolf}, \citenamefont {Nagel},
  \citenamefont {Meingast},\ and\ \citenamefont {Schuppler}}]{Merz2021}%
  \BibitemOpen
  \bibfield  {author} {\bibinfo {author} {\bibfnamefont {M.}~\bibnamefont
  {Merz}}, \bibinfo {author} {\bibfnamefont {L.}~\bibnamefont {Wang}}, \bibinfo
  {author} {\bibfnamefont {T.}~\bibnamefont {Wolf}}, \bibinfo {author}
  {\bibfnamefont {P.}~\bibnamefont {Nagel}}, \bibinfo {author} {\bibfnamefont
  {C.}~\bibnamefont {Meingast}}, \ and\ \bibinfo {author} {\bibfnamefont
  {S.}~\bibnamefont {Schuppler}},\ }\href {\doibase
  10.1103/physrevb.104.184509} {\bibfield  {journal} {\bibinfo  {journal}
  {Physical Review B}\ }\textbf {\bibinfo {volume} {104}},\ \bibinfo {pages}
  {184509} (\bibinfo {year} {2021})}\BibitemShut {NoStop}%
\bibitem [{\citenamefont {Chu}\ \emph {et~al.}(2010)\citenamefont {Chu},
  \citenamefont {Analytis}, \citenamefont {Greve}, \citenamefont {McMahon},
  \citenamefont {Islam}, \citenamefont {Yamamoto},\ and\ \citenamefont
  {Fisher}}]{Chu2010}%
  \BibitemOpen
  \bibfield  {author} {\bibinfo {author} {\bibfnamefont {J.-H.}\ \bibnamefont
  {Chu}}, \bibinfo {author} {\bibfnamefont {J.~G.}\ \bibnamefont {Analytis}},
  \bibinfo {author} {\bibfnamefont {K.~D.}\ \bibnamefont {Greve}}, \bibinfo
  {author} {\bibfnamefont {P.~L.}\ \bibnamefont {McMahon}}, \bibinfo {author}
  {\bibfnamefont {Z.}~\bibnamefont {Islam}}, \bibinfo {author} {\bibfnamefont
  {Y.}~\bibnamefont {Yamamoto}}, \ and\ \bibinfo {author} {\bibfnamefont
  {I.~R.}\ \bibnamefont {Fisher}},\ }\href {\doibase 10.1126/science.1190482}
  {\bibfield  {journal} {\bibinfo  {journal} {Science}\ }\textbf {\bibinfo
  {volume} {329}},\ \bibinfo {pages} {824} (\bibinfo {year}
  {2010})}\BibitemShut {NoStop}%
\bibitem [{\citenamefont {Ronning}\ \emph {et~al.}(2008)\citenamefont
  {Ronning}, \citenamefont {Kurita}, \citenamefont {Bauer}, \citenamefont
  {Scott}, \citenamefont {Park}, \citenamefont {Klimczuk}, \citenamefont
  {Movshovich},\ and\ \citenamefont {Thompson}}]{Ronning2008}%
  \BibitemOpen
  \bibfield  {author} {\bibinfo {author} {\bibfnamefont {F.}~\bibnamefont
  {Ronning}}, \bibinfo {author} {\bibfnamefont {N.}~\bibnamefont {Kurita}},
  \bibinfo {author} {\bibfnamefont {E.~D.}\ \bibnamefont {Bauer}}, \bibinfo
  {author} {\bibfnamefont {B.~L.}\ \bibnamefont {Scott}}, \bibinfo {author}
  {\bibfnamefont {T.}~\bibnamefont {Park}}, \bibinfo {author} {\bibfnamefont
  {T.}~\bibnamefont {Klimczuk}}, \bibinfo {author} {\bibfnamefont
  {R.}~\bibnamefont {Movshovich}}, \ and\ \bibinfo {author} {\bibfnamefont
  {J.~D.}\ \bibnamefont {Thompson}},\ }\href {\doibase
  10.1088/0953-8984/20/34/342203} {\bibfield  {journal} {\bibinfo  {journal}
  {Journal of Physics: Condensed Matter}\ }\textbf {\bibinfo {volume} {20}},\
  \bibinfo {pages} {342203} (\bibinfo {year} {2008})}\BibitemShut {NoStop}%
\bibitem [{\citenamefont {Sefat}\ \emph {et~al.}(2009)\citenamefont {Sefat},
  \citenamefont {McGuire}, \citenamefont {Jin}, \citenamefont {Sales},
  \citenamefont {Mandrus}, \citenamefont {Ronning}, \citenamefont {Bauer},\
  and\ \citenamefont {Mozharivskyj}}]{Sefat2009}%
  \BibitemOpen
  \bibfield  {author} {\bibinfo {author} {\bibfnamefont {A.~S.}\ \bibnamefont
  {Sefat}}, \bibinfo {author} {\bibfnamefont {M.~A.}\ \bibnamefont {McGuire}},
  \bibinfo {author} {\bibfnamefont {R.}~\bibnamefont {Jin}}, \bibinfo {author}
  {\bibfnamefont {B.~C.}\ \bibnamefont {Sales}}, \bibinfo {author}
  {\bibfnamefont {D.}~\bibnamefont {Mandrus}}, \bibinfo {author} {\bibfnamefont
  {F.}~\bibnamefont {Ronning}}, \bibinfo {author} {\bibfnamefont {E.~D.}\
  \bibnamefont {Bauer}}, \ and\ \bibinfo {author} {\bibfnamefont
  {Y.}~\bibnamefont {Mozharivskyj}},\ }\href {\doibase
  10.1103/physrevb.79.094508} {\bibfield  {journal} {\bibinfo  {journal}
  {Physical Review B}\ }\textbf {\bibinfo {volume} {79}},\ \bibinfo {pages}
  {094508} (\bibinfo {year} {2009})}\BibitemShut {NoStop}%
\bibitem [{\citenamefont {Kudo}\ \emph {et~al.}(2012)\citenamefont {Kudo},
  \citenamefont {Takasuga}, \citenamefont {Okamoto}, \citenamefont {Hiroi},\
  and\ \citenamefont {Nohara}}]{Kudo2012}%
  \BibitemOpen
  \bibfield  {author} {\bibinfo {author} {\bibfnamefont {K.}~\bibnamefont
  {Kudo}}, \bibinfo {author} {\bibfnamefont {M.}~\bibnamefont {Takasuga}},
  \bibinfo {author} {\bibfnamefont {Y.}~\bibnamefont {Okamoto}}, \bibinfo
  {author} {\bibfnamefont {Z.}~\bibnamefont {Hiroi}}, \ and\ \bibinfo {author}
  {\bibfnamefont {M.}~\bibnamefont {Nohara}},\ }\href {\doibase
  10.1103/physrevlett.109.097002} {\bibfield  {journal} {\bibinfo  {journal}
  {Physical Review Letters}\ }\textbf {\bibinfo {volume} {109}},\ \bibinfo
  {pages} {097002} (\bibinfo {year} {2012})}\BibitemShut {NoStop}%
\bibitem [{\citenamefont {Kurita}\ \emph {et~al.}(2009)\citenamefont {Kurita},
  \citenamefont {Ronning}, \citenamefont {Tokiwa}, \citenamefont {Bauer},
  \citenamefont {Subedi}, \citenamefont {Singh}, \citenamefont {Thompson},\
  and\ \citenamefont {Movshovich}}]{Kurita2009}%
  \BibitemOpen
  \bibfield  {author} {\bibinfo {author} {\bibfnamefont {N.}~\bibnamefont
  {Kurita}}, \bibinfo {author} {\bibfnamefont {F.}~\bibnamefont {Ronning}},
  \bibinfo {author} {\bibfnamefont {Y.}~\bibnamefont {Tokiwa}}, \bibinfo
  {author} {\bibfnamefont {E.~D.}\ \bibnamefont {Bauer}}, \bibinfo {author}
  {\bibfnamefont {A.}~\bibnamefont {Subedi}}, \bibinfo {author} {\bibfnamefont
  {D.~J.}\ \bibnamefont {Singh}}, \bibinfo {author} {\bibfnamefont {J.~D.}\
  \bibnamefont {Thompson}}, \ and\ \bibinfo {author} {\bibfnamefont
  {R.}~\bibnamefont {Movshovich}},\ }\href {\doibase
  10.1103/physrevlett.102.147004} {\bibfield  {journal} {\bibinfo  {journal}
  {Physical Review Letters}\ }\textbf {\bibinfo {volume} {102}},\ \bibinfo
  {pages} {147004} (\bibinfo {year} {2009})}\BibitemShut {NoStop}%
\bibitem [{SOM()}]{SOM}%
  \BibitemOpen
  \href@noop {} {}\bibinfo {note} {See Supplemental Material at XXX for
  information on the samples, the experimental and computational methods as
  well as on the data analysis, which includes also
  references~\cite{Sheldrick2008,vaklav2014,louie79,meyer97,haman79,bache82,vande85,perde96,heid99,Feng2012,Currat2006}.}\BibitemShut
  {Stop}%
\bibitem [{\citenamefont {Pokharel}\ \emph {et~al.}(2022)\citenamefont
  {Pokharel}, \citenamefont {Grigorev}, \citenamefont {Mejas}, \citenamefont
  {Dong}, \citenamefont {Haghighirad}, \citenamefont {Heid}, \citenamefont
  {Yao}, \citenamefont {Merz}, \citenamefont {Le~Tacon},\ and\ \citenamefont
  {Demsar}}]{Grigorev2021}%
  \BibitemOpen
  \bibfield  {author} {\bibinfo {author} {\bibfnamefont {A.~R.}\ \bibnamefont
  {Pokharel}}, \bibinfo {author} {\bibfnamefont {V.}~\bibnamefont {Grigorev}},
  \bibinfo {author} {\bibfnamefont {A.}~\bibnamefont {Mejas}}, \bibinfo
  {author} {\bibfnamefont {T.}~\bibnamefont {Dong}}, \bibinfo {author}
  {\bibfnamefont {A.~A.}\ \bibnamefont {Haghighirad}}, \bibinfo {author}
  {\bibfnamefont {R.}~\bibnamefont {Heid}}, \bibinfo {author} {\bibfnamefont
  {Y.}~\bibnamefont {Yao}}, \bibinfo {author} {\bibfnamefont {M.}~\bibnamefont
  {Merz}}, \bibinfo {author} {\bibfnamefont {M.}~\bibnamefont {Le~Tacon}}, \
  and\ \bibinfo {author} {\bibfnamefont {J.}~\bibnamefont {Demsar}},\ }\href
  {\doibase 10.1038/s42005-022-00919-x} {\bibfield  {journal} {\bibinfo
  {journal} {Communications Physics}\ }\textbf {\bibinfo {volume} {5}},\
  \bibinfo {pages} {141} (\bibinfo {year} {2022})}\BibitemShut {NoStop}%
\bibitem [{\citenamefont {Meingast}\ \emph {et~al.}(2022)\citenamefont
  {Meingast}, \citenamefont {Shukla}, \citenamefont {Wang}, \citenamefont
  {Heid}, \citenamefont {Hardy}, \citenamefont {Frachet}, \citenamefont
  {Willa}, \citenamefont {Lacmann}, \citenamefont {Le~Tacon}, \citenamefont
  {Merz}, \citenamefont {Haghighirad},\ and\ \citenamefont
  {Wolf}}]{Meingast2022}%
  \BibitemOpen
  \bibfield  {author} {\bibinfo {author} {\bibfnamefont {C.}~\bibnamefont
  {Meingast}}, \bibinfo {author} {\bibfnamefont {A.}~\bibnamefont {Shukla}},
  \bibinfo {author} {\bibfnamefont {L.}~\bibnamefont {Wang}}, \bibinfo {author}
  {\bibfnamefont {R.}~\bibnamefont {Heid}}, \bibinfo {author} {\bibfnamefont
  {F.}~\bibnamefont {Hardy}}, \bibinfo {author} {\bibfnamefont
  {M.}~\bibnamefont {Frachet}}, \bibinfo {author} {\bibfnamefont
  {K.}~\bibnamefont {Willa}}, \bibinfo {author} {\bibfnamefont
  {T.}~\bibnamefont {Lacmann}}, \bibinfo {author} {\bibfnamefont
  {M.}~\bibnamefont {Le~Tacon}}, \bibinfo {author} {\bibfnamefont
  {M.}~\bibnamefont {Merz}}, \bibinfo {author} {\bibfnamefont {A.-A.}\
  \bibnamefont {Haghighirad}}, \ and\ \bibinfo {author} {\bibfnamefont
  {T.}~\bibnamefont {Wolf}},\ }\href {\doibase 10.1103/PhysRevB.106.144507}
  {\bibfield  {journal} {\bibinfo  {journal} {Physical Review B}\ }\textbf
  {\bibinfo {volume} {106}},\ \bibinfo {pages} {144507} (\bibinfo {year}
  {2022})}\BibitemShut {NoStop}%
\bibitem [{\citenamefont {Frachet}\ \emph {et~al.}(2022)\citenamefont
  {Frachet}, \citenamefont {Wiecki}, \citenamefont {Lacmann}, \citenamefont
  {Souliou}, \citenamefont {Willa}, \citenamefont {Meingast}, \citenamefont
  {Merz}, \citenamefont {Haghighirad}, \citenamefont {Le~Tacon},\ and\
  \citenamefont {B\"{o}hmer}}]{Frachet2022}%
  \BibitemOpen
  \bibfield  {author} {\bibinfo {author} {\bibfnamefont {M.}~\bibnamefont
  {Frachet}}, \bibinfo {author} {\bibfnamefont {P.~W.}\ \bibnamefont {Wiecki}},
  \bibinfo {author} {\bibfnamefont {T.}~\bibnamefont {Lacmann}}, \bibinfo
  {author} {\bibfnamefont {S.-M.}\ \bibnamefont {Souliou}}, \bibinfo {author}
  {\bibfnamefont {K.}~\bibnamefont {Willa}}, \bibinfo {author} {\bibfnamefont
  {C.}~\bibnamefont {Meingast}}, \bibinfo {author} {\bibfnamefont
  {M.}~\bibnamefont {Merz}}, \bibinfo {author} {\bibfnamefont {A.-A.}\
  \bibnamefont {Haghighirad}}, \bibinfo {author} {\bibfnamefont
  {M.}~\bibnamefont {Le~Tacon}}, \ and\ \bibinfo {author} {\bibfnamefont
  {A.~E.}\ \bibnamefont {B\"{o}hmer}},\ }\href@noop {} {\  (\bibinfo {year}
  {2022})},\ \Eprint {http://arxiv.org/abs/2207.02462} {arXiv:2207.02462}
  \BibitemShut {NoStop}%
\bibitem [{\citenamefont {Subedi}\ and\ \citenamefont
  {Singh}(2008)}]{Subedi2008}%
  \BibitemOpen
  \bibfield  {author} {\bibinfo {author} {\bibfnamefont {A.}~\bibnamefont
  {Subedi}}\ and\ \bibinfo {author} {\bibfnamefont {D.~J.}\ \bibnamefont
  {Singh}},\ }\href {\doibase 10.1103/physrevb.78.132511} {\bibfield  {journal}
  {\bibinfo  {journal} {Physical Review B}\ }\textbf {\bibinfo {volume} {78}},\
  \bibinfo {pages} {132511} (\bibinfo {year} {2008})}\BibitemShut {NoStop}%
\bibitem [{\citenamefont {Yao}\ \emph {et~al.}(2022)\citenamefont {Yao},
  \citenamefont {Willa}, \citenamefont {Lacmann}, \citenamefont {Souliou},
  \citenamefont {Frachet}, \citenamefont {Willa}, \citenamefont {Merz},
  \citenamefont {Weber}, \citenamefont {Meingast}, \citenamefont {Heid},
  \citenamefont {Haghighirad}, \citenamefont {Schmalian},\ and\ \citenamefont
  {Le~Tacon}}]{Yao2022}%
  \BibitemOpen
  \bibfield  {author} {\bibinfo {author} {\bibfnamefont {Y.}~\bibnamefont
  {Yao}}, \bibinfo {author} {\bibfnamefont {R.}~\bibnamefont {Willa}}, \bibinfo
  {author} {\bibfnamefont {T.}~\bibnamefont {Lacmann}}, \bibinfo {author}
  {\bibfnamefont {S.-M.}\ \bibnamefont {Souliou}}, \bibinfo {author}
  {\bibfnamefont {M.}~\bibnamefont {Frachet}}, \bibinfo {author} {\bibfnamefont
  {K.}~\bibnamefont {Willa}}, \bibinfo {author} {\bibfnamefont
  {M.}~\bibnamefont {Merz}}, \bibinfo {author} {\bibfnamefont {F.}~\bibnamefont
  {Weber}}, \bibinfo {author} {\bibfnamefont {C.}~\bibnamefont {Meingast}},
  \bibinfo {author} {\bibfnamefont {R.}~\bibnamefont {Heid}}, \bibinfo {author}
  {\bibfnamefont {A.-A.}\ \bibnamefont {Haghighirad}}, \bibinfo {author}
  {\bibfnamefont {J.}~\bibnamefont {Schmalian}}, \ and\ \bibinfo {author}
  {\bibfnamefont {M.}~\bibnamefont {Le~Tacon}},\ }\href {\doibase
  10.1038/s41467-022-32112-7} {\bibfield  {journal} {\bibinfo  {journal}
  {Nature Communications}\ }\textbf {\bibinfo {volume} {13}},\ \bibinfo {pages}
  {4535} (\bibinfo {year} {2022})}\BibitemShut {NoStop}%
\bibitem{Kim_Science2018} H. H. Kim, S. M. Souliou, M. E. Barber, E. Lefran\c cois, M. Minola, M. Tortora, R. Heid, N. Nandi, R. A. Borzi, G. Garbarino, A. Bosak, J. Porras, T. Loew, M. K\"{o}nig, P. M. Moll, A. P. Mackenzie, B. Keimer, C. W. Hicks, and M. Le Tacon, \href{http://science.sciencemag.org/content/362/6418/1040.abstract}{Science {\bf 362}, 1040 (2018)}.
\bibitem [{\citenamefont {Guo}\ \emph {et~al.}(2022)\citenamefont {Guo},
  \citenamefont {Klemm}, \citenamefont {Oh}, \citenamefont {Xie}, \citenamefont
  {Lei}, \citenamefont {Gorovikov}, \citenamefont {Pedersen}, \citenamefont
  {Michiardi}, \citenamefont {Zhdanovich}, \citenamefont {Damascelli},
  \citenamefont {Denlinger}, \citenamefont {Hashimoto}, \citenamefont {Lu},
  \citenamefont {Mo}, \citenamefont {Moore}, \citenamefont {Birgeneau},
  \citenamefont {Singh}, \citenamefont {Dai},\ and\ \citenamefont
  {Yi}}]{Guo2022}%
  \BibitemOpen
  \bibfield  {author} {\bibinfo {author} {\bibfnamefont {Y.}~\bibnamefont
  {Guo}}, \bibinfo {author} {\bibfnamefont {M.}~\bibnamefont {Klemm}}, \bibinfo
  {author} {\bibfnamefont {J.~S.}\ \bibnamefont {Oh}}, \bibinfo {author}
  {\bibfnamefont {Y.}~\bibnamefont {Xie}}, \bibinfo {author} {\bibfnamefont
  {B.-H.}\ \bibnamefont {Lei}}, \bibinfo {author} {\bibfnamefont
  {S.}~\bibnamefont {Gorovikov}}, \bibinfo {author} {\bibfnamefont
  {T.}~\bibnamefont {Pedersen}}, \bibinfo {author} {\bibfnamefont
  {M.}~\bibnamefont {Michiardi}}, \bibinfo {author} {\bibfnamefont
  {S.}~\bibnamefont {Zhdanovich}}, \bibinfo {author} {\bibfnamefont
  {A.}~\bibnamefont {Damascelli}}, \bibinfo {author} {\bibfnamefont
  {J.}~\bibnamefont {Denlinger}}, \bibinfo {author} {\bibfnamefont
  {M.}~\bibnamefont {Hashimoto}}, \bibinfo {author} {\bibfnamefont
  {D.}~\bibnamefont {Lu}}, \bibinfo {author} {\bibfnamefont {S.-K.}\
  \bibnamefont {Mo}}, \bibinfo {author} {\bibfnamefont {R.~G.}\ \bibnamefont
  {Moore}}, \bibinfo {author} {\bibfnamefont {R.~J.}\ \bibnamefont
  {Birgeneau}}, \bibinfo {author} {\bibfnamefont {D.~J.}\ \bibnamefont
  {Singh}}, \bibinfo {author} {\bibfnamefont {P.}~\bibnamefont {Dai}}, \ and\
  \bibinfo {author} {\bibfnamefont {M.}~\bibnamefont {Yi}},\ }\href@noop {} {\
  (\bibinfo {year} {2022})},\ \Eprint {http://arxiv.org/abs/2205.14339}
  {arXiv:2205.14339} \BibitemShut {NoStop}%
\bibitem [{\citenamefont {Pavlov}\ \emph {et~al.}(2021)\citenamefont {Pavlov},
  \citenamefont {Kim}, \citenamefont {Yaresko}, \citenamefont {Choi},
  \citenamefont {Nekrasov},\ and\ \citenamefont {Evtushinsky}}]{Pavlov2021}%
  \BibitemOpen
  \bibfield  {author} {\bibinfo {author} {\bibfnamefont {N.~S.}\ \bibnamefont
  {Pavlov}}, \bibinfo {author} {\bibfnamefont {T.~K.}\ \bibnamefont {Kim}},
  \bibinfo {author} {\bibfnamefont {A.}~\bibnamefont {Yaresko}}, \bibinfo
  {author} {\bibfnamefont {K.-Y.}\ \bibnamefont {Choi}}, \bibinfo {author}
  {\bibfnamefont {I.~A.}\ \bibnamefont {Nekrasov}}, \ and\ \bibinfo {author}
  {\bibfnamefont {D.~V.}\ \bibnamefont {Evtushinsky}},\ }\href {\doibase
  10.1021/acs.jpcc.1c08142} {\bibfield  {journal} {\bibinfo  {journal} {The
  Journal of Physical Chemistry C}\ }\textbf {\bibinfo {volume} {125}},\
  \bibinfo {pages} {28075} (\bibinfo {year} {2021})}\BibitemShut {NoStop}%
\bibitem [{\citenamefont {Girard}\ \emph {et~al.}(2019)\citenamefont {Girard},
  \citenamefont {Nguyen-Thanh}, \citenamefont {Souliou}, \citenamefont
  {Stekiel}, \citenamefont {Morgenroth}, \citenamefont {Paolasini},
  \citenamefont {Minelli}, \citenamefont {Gambetti}, \citenamefont {Winkler},\
  and\ \citenamefont {Bosak}}]{Girard2019}%
  \BibitemOpen
  \bibfield  {author} {\bibinfo {author} {\bibfnamefont {A.}~\bibnamefont
  {Girard}}, \bibinfo {author} {\bibfnamefont {T.}~\bibnamefont
  {Nguyen-Thanh}}, \bibinfo {author} {\bibfnamefont {S.~M.}\ \bibnamefont
  {Souliou}}, \bibinfo {author} {\bibfnamefont {M.}~\bibnamefont {Stekiel}},
  \bibinfo {author} {\bibfnamefont {W.}~\bibnamefont {Morgenroth}}, \bibinfo
  {author} {\bibfnamefont {L.}~\bibnamefont {Paolasini}}, \bibinfo {author}
  {\bibfnamefont {A.}~\bibnamefont {Minelli}}, \bibinfo {author} {\bibfnamefont
  {D.}~\bibnamefont {Gambetti}}, \bibinfo {author} {\bibfnamefont
  {B.}~\bibnamefont {Winkler}}, \ and\ \bibinfo {author} {\bibfnamefont
  {A.}~\bibnamefont {Bosak}},\ }\href {\doibase 10.1107/S1600577518016132}
  {\bibfield  {journal} {\bibinfo  {journal} {Journal of Synchrotron
  Radiation}\ }\textbf {\bibinfo {volume} {26}},\ \bibinfo {pages} {272}
  (\bibinfo {year} {2019})}\BibitemShut {NoStop}%
\bibitem [{\citenamefont {Krisch}\ and\ \citenamefont {Sette}()}]{Krisch2006}%
  \BibitemOpen
  \bibfield  {author} {\bibinfo {author} {\bibfnamefont {M.}~\bibnamefont
  {Krisch}}\ and\ \bibinfo {author} {\bibfnamefont {F.}~\bibnamefont {Sette}}\
  }(\bibinfo  {publisher} {Springer Berlin Heidelberg})\ pp.\ \bibinfo {pages}
  {317--370}\BibitemShut {NoStop}%
\bibitem [{\citenamefont {Song}\ \emph {et~al.}(2022)\citenamefont {Song},
  \citenamefont {Wu}, \citenamefont {Chen}, \citenamefont {He}, \citenamefont
  {Uchiyama}, \citenamefont {Li}, \citenamefont {Cao}, \citenamefont {Guo},
  \citenamefont {Cao},\ and\ \citenamefont {Birgeneau}}]{Song2022}%
  \BibitemOpen
  \bibfield  {author} {\bibinfo {author} {\bibfnamefont {Y.}~\bibnamefont
  {Song}}, \bibinfo {author} {\bibfnamefont {S.}~\bibnamefont {Wu}}, \bibinfo
  {author} {\bibfnamefont {X.}~\bibnamefont {Chen}}, \bibinfo {author}
  {\bibfnamefont {Y.}~\bibnamefont {He}}, \bibinfo {author} {\bibfnamefont
  {H.}~\bibnamefont {Uchiyama}}, \bibinfo {author} {\bibfnamefont
  {B.}~\bibnamefont {Li}}, \bibinfo {author} {\bibfnamefont {S.}~\bibnamefont
  {Cao}}, \bibinfo {author} {\bibfnamefont {J.}~\bibnamefont {Guo}}, \bibinfo
  {author} {\bibfnamefont {G.}~\bibnamefont {Cao}}, \ and\ \bibinfo {author}
  {\bibfnamefont {R.}~\bibnamefont {Birgeneau}},\ }\href@noop {} {\  (\bibinfo
  {year} {2022})},\ \Eprint {http://arxiv.org/abs/2207.03289}
  {arXiv:2207.03289} \BibitemShut {NoStop}%
\bibitem [{\citenamefont {Zhou}\ \emph {et~al.}(2011)\citenamefont {Zhou},
  \citenamefont {Xu}, \citenamefont {Zhang}, \citenamefont {Xu}, \citenamefont
  {He}, \citenamefont {Yang}, \citenamefont {Chen}, \citenamefont {Xie},
  \citenamefont {Cui}, \citenamefont {Arita}, \citenamefont {Shimada},
  \citenamefont {Namatame}, \citenamefont {Taniguchi}, \citenamefont {Dai},\
  and\ \citenamefont {Feng}}]{Zhou2011}%
  \BibitemOpen
  \bibfield  {author} {\bibinfo {author} {\bibfnamefont {B.}~\bibnamefont
  {Zhou}}, \bibinfo {author} {\bibfnamefont {M.}~\bibnamefont {Xu}}, \bibinfo
  {author} {\bibfnamefont {Y.}~\bibnamefont {Zhang}}, \bibinfo {author}
  {\bibfnamefont {G.}~\bibnamefont {Xu}}, \bibinfo {author} {\bibfnamefont
  {C.}~\bibnamefont {He}}, \bibinfo {author} {\bibfnamefont {L.~X.}\
  \bibnamefont {Yang}}, \bibinfo {author} {\bibfnamefont {F.}~\bibnamefont
  {Chen}}, \bibinfo {author} {\bibfnamefont {B.~P.}\ \bibnamefont {Xie}},
  \bibinfo {author} {\bibfnamefont {X.-Y.}\ \bibnamefont {Cui}}, \bibinfo
  {author} {\bibfnamefont {M.}~\bibnamefont {Arita}}, \bibinfo {author}
  {\bibfnamefont {K.}~\bibnamefont {Shimada}}, \bibinfo {author} {\bibfnamefont
  {H.}~\bibnamefont {Namatame}}, \bibinfo {author} {\bibfnamefont
  {M.}~\bibnamefont {Taniguchi}}, \bibinfo {author} {\bibfnamefont
  {X.}~\bibnamefont {Dai}}, \ and\ \bibinfo {author} {\bibfnamefont {D.~L.}\
  \bibnamefont {Feng}},\ }\href {\doibase 10.1103/physrevb.83.035110}
  {\bibfield  {journal} {\bibinfo  {journal} {Physical Review B}\ }\textbf
  {\bibinfo {volume} {83}},\ \bibinfo {pages} {035110} (\bibinfo {year}
  {2011})}\BibitemShut {NoStop}%
\bibitem [{\citenamefont {Peierls}(2001)}]{Peierls2001}%
  \BibitemOpen
  \bibfield  {author} {\bibinfo {author} {\bibfnamefont {R.~E.}\ \bibnamefont
  {Peierls}},\ }\href {\doibase 10.1093/acprof:oso/9780198507819.001.0001}
  {\emph {\bibinfo {title} {Quantum Theory of Solids}}}\ (\bibinfo  {publisher}
  {Oxford University Press},\ \bibinfo {year} {2001})\BibitemShut {NoStop}%
\bibitem [{\citenamefont {Gr\"{u}ner}(2018)}]{Gruener2018}%
  \BibitemOpen
  \bibfield  {author} {\bibinfo {author} {\bibfnamefont {G.}~\bibnamefont
  {Gr\"{u}ner}},\ }\href {\doibase 10.1201/9780429501012} {\emph {\bibinfo
  {title} {Density Waves in Solids}}}\ (\bibinfo  {publisher} {{CRC} Press},\
  \bibinfo {year} {2018})\BibitemShut {NoStop}%
\bibitem [{\citenamefont {Weber}\ \emph
  {et~al.}(2011{\natexlab{a}})\citenamefont {Weber}, \citenamefont
  {Rosenkranz}, \citenamefont {Castellan}, \citenamefont {Osborn},
  \citenamefont {Hott}, \citenamefont {Heid}, \citenamefont {Bohnen},
  \citenamefont {Egami}, \citenamefont {Said},\ and\ \citenamefont
  {Reznik}}]{Weber2011}%
  \BibitemOpen
  \bibfield  {author} {\bibinfo {author} {\bibfnamefont {F.}~\bibnamefont
  {Weber}}, \bibinfo {author} {\bibfnamefont {S.}~\bibnamefont {Rosenkranz}},
  \bibinfo {author} {\bibfnamefont {J.-P.}\ \bibnamefont {Castellan}}, \bibinfo
  {author} {\bibfnamefont {R.}~\bibnamefont {Osborn}}, \bibinfo {author}
  {\bibfnamefont {R.}~\bibnamefont {Hott}}, \bibinfo {author} {\bibfnamefont
  {R.}~\bibnamefont {Heid}}, \bibinfo {author} {\bibfnamefont {K.-P.}\
  \bibnamefont {Bohnen}}, \bibinfo {author} {\bibfnamefont {T.}~\bibnamefont
  {Egami}}, \bibinfo {author} {\bibfnamefont {A.~H.}\ \bibnamefont {Said}}, \
  and\ \bibinfo {author} {\bibfnamefont {D.}~\bibnamefont {Reznik}},\ }\href
  {\doibase 10.1103/physrevlett.107.107403} {\bibfield  {journal} {\bibinfo
  {journal} {Physical Review Letters}\ }\textbf {\bibinfo {volume} {107}},\
  \bibinfo {pages} {107403} (\bibinfo {year} {2011}{\natexlab{a}})}\BibitemShut
  {NoStop}%
\bibitem [{\citenamefont {Weber}\ \emph
  {et~al.}(2011{\natexlab{b}})\citenamefont {Weber}, \citenamefont
  {Rosenkranz}, \citenamefont {Castellan}, \citenamefont {Osborn},
  \citenamefont {Karapetrov}, \citenamefont {Hott}, \citenamefont {Heid},
  \citenamefont {Bohnen},\ and\ \citenamefont {Alatas}}]{Weber2011a}%
  \BibitemOpen
  \bibfield  {author} {\bibinfo {author} {\bibfnamefont {F.}~\bibnamefont
  {Weber}}, \bibinfo {author} {\bibfnamefont {S.}~\bibnamefont {Rosenkranz}},
  \bibinfo {author} {\bibfnamefont {J.-P.}\ \bibnamefont {Castellan}}, \bibinfo
  {author} {\bibfnamefont {R.}~\bibnamefont {Osborn}}, \bibinfo {author}
  {\bibfnamefont {G.}~\bibnamefont {Karapetrov}}, \bibinfo {author}
  {\bibfnamefont {R.}~\bibnamefont {Hott}}, \bibinfo {author} {\bibfnamefont
  {R.}~\bibnamefont {Heid}}, \bibinfo {author} {\bibfnamefont {K.-P.}\
  \bibnamefont {Bohnen}}, \ and\ \bibinfo {author} {\bibfnamefont
  {A.}~\bibnamefont {Alatas}},\ }\href {\doibase
  10.1103/physrevlett.107.266401} {\bibfield  {journal} {\bibinfo  {journal}
  {Physical Review Letters}\ }\textbf {\bibinfo {volume} {107}},\ \bibinfo
  {pages} {266401} (\bibinfo {year} {2011}{\natexlab{b}})}\BibitemShut
  {NoStop}%
\bibitem [{\citenamefont {Bosak}\ \emph {et~al.}(2021)\citenamefont {Bosak},
  \citenamefont {Souliou}, \citenamefont {Faugeras}, \citenamefont {Heid},
  \citenamefont {Molas}, \citenamefont {Chen}, \citenamefont {Wang},
  \citenamefont {Potemski},\ and\ \citenamefont {Le~Tacon}}]{Bosak2021}%
  \BibitemOpen
  \bibfield  {author} {\bibinfo {author} {\bibfnamefont {A.}~\bibnamefont
  {Bosak}}, \bibinfo {author} {\bibfnamefont {S.-M.}\ \bibnamefont {Souliou}},
  \bibinfo {author} {\bibfnamefont {C.}~\bibnamefont {Faugeras}}, \bibinfo
  {author} {\bibfnamefont {R.}~\bibnamefont {Heid}}, \bibinfo {author}
  {\bibfnamefont {M.~R.}\ \bibnamefont {Molas}}, \bibinfo {author}
  {\bibfnamefont {R.-Y.}\ \bibnamefont {Chen}}, \bibinfo {author}
  {\bibfnamefont {N.-L.}\ \bibnamefont {Wang}}, \bibinfo {author}
  {\bibfnamefont {M.}~\bibnamefont {Potemski}}, \ and\ \bibinfo {author}
  {\bibfnamefont {M.}~\bibnamefont {Le~Tacon}},\ }\href {\doibase
  10.1103/physrevresearch.3.033020} {\bibfield  {journal} {\bibinfo  {journal}
  {Physical Review Research}\ }\textbf {\bibinfo {volume} {3}},\ \bibinfo
  {pages} {033020} (\bibinfo {year} {2021})}\BibitemShut {NoStop}%
\bibitem [{\citenamefont {Noda}\ \emph {et~al.}(2017)\citenamefont {Noda},
  \citenamefont {Kudo}, \citenamefont {Takasuga}, \citenamefont {Nohara},
  \citenamefont {Sugimoto}, \citenamefont {Ootsuki}, \citenamefont {Kobayashi},
  \citenamefont {Horiba}, \citenamefont {Ono}, \citenamefont {Kumigashira},
  \citenamefont {Fujimori}, \citenamefont {Saini},\ and\ \citenamefont
  {Mizokawa}}]{Noda2017}%
  \BibitemOpen
  \bibfield  {author} {\bibinfo {author} {\bibfnamefont {T.}~\bibnamefont
  {Noda}}, \bibinfo {author} {\bibfnamefont {K.}~\bibnamefont {Kudo}}, \bibinfo
  {author} {\bibfnamefont {M.}~\bibnamefont {Takasuga}}, \bibinfo {author}
  {\bibfnamefont {M.}~\bibnamefont {Nohara}}, \bibinfo {author} {\bibfnamefont
  {T.}~\bibnamefont {Sugimoto}}, \bibinfo {author} {\bibfnamefont
  {D.}~\bibnamefont {Ootsuki}}, \bibinfo {author} {\bibfnamefont
  {M.}~\bibnamefont {Kobayashi}}, \bibinfo {author} {\bibfnamefont
  {K.}~\bibnamefont {Horiba}}, \bibinfo {author} {\bibfnamefont
  {K.}~\bibnamefont {Ono}}, \bibinfo {author} {\bibfnamefont {H.}~\bibnamefont
  {Kumigashira}}, \bibinfo {author} {\bibfnamefont {A.}~\bibnamefont
  {Fujimori}}, \bibinfo {author} {\bibfnamefont {N.~L.}\ \bibnamefont {Saini}},
  \ and\ \bibinfo {author} {\bibfnamefont {T.}~\bibnamefont {Mizokawa}},\
  }\href {\doibase 10.7566/jpsj.86.064708} {\bibfield  {journal} {\bibinfo
  {journal} {Journal of the Physical Society of Japan}\ }\textbf {\bibinfo
  {volume} {86}},\ \bibinfo {pages} {064708} (\bibinfo {year}
  {2017})}\BibitemShut {NoStop}%
\bibitem [{\citenamefont {Flicker}\ and\ \citenamefont {van
  Wezel}(2015)}]{Flicker_NatCom2015}%
  \BibitemOpen
  \bibfield  {author} {\bibinfo {author} {\bibfnamefont {F.}~\bibnamefont
  {Flicker}}\ and\ \bibinfo {author} {\bibfnamefont {J.}~\bibnamefont {van
  Wezel}},\ }\href {\doibase 10.1038/ncomms8034} {\bibfield  {journal}
  {\bibinfo  {journal} {Nature Communications}\ }\textbf {\bibinfo {volume}
  {6}},\ \bibinfo {pages} {7034} (\bibinfo {year} {2015})}\BibitemShut
  {NoStop}%
\bibitem [{\citenamefont {Khomskii}\ and\ \citenamefont
  {Mizokawa}(2005)}]{Khomskii2005}%
  \BibitemOpen
  \bibfield  {author} {\bibinfo {author} {\bibfnamefont {D.~I.}\ \bibnamefont
  {Khomskii}}\ and\ \bibinfo {author} {\bibfnamefont {T.}~\bibnamefont
  {Mizokawa}},\ }\href {\doibase 10.1103/physrevlett.94.156402} {\bibfield
  {journal} {\bibinfo  {journal} {Physical Review Letters}\ }\textbf {\bibinfo
  {volume} {94}},\ \bibinfo {pages} {156402} (\bibinfo {year}
  {2005})}\BibitemShut {NoStop}%
\bibitem [{\citenamefont {Streltsov}\ and\ \citenamefont
  {Khomskii}(2014)}]{Streltsov2014}%
  \BibitemOpen
  \bibfield  {author} {\bibinfo {author} {\bibfnamefont {S.~V.}\ \bibnamefont
  {Streltsov}}\ and\ \bibinfo {author} {\bibfnamefont {D.~I.}\ \bibnamefont
  {Khomskii}},\ }\href {\doibase 10.1103/physrevb.89.161112} {\bibfield
  {journal} {\bibinfo  {journal} {Physical Review B}\ }\textbf {\bibinfo
  {volume} {89}},\ \bibinfo {pages} {161112} (\bibinfo {year}
  {2014})}\BibitemShut {NoStop}%
\bibitem [{\citenamefont {Khomskii}\ and\ \citenamefont
  {Streltsov}(2020)}]{Khomskii2020}%
  \BibitemOpen
  \bibfield  {author} {\bibinfo {author} {\bibfnamefont {D.~I.}\ \bibnamefont
  {Khomskii}}\ and\ \bibinfo {author} {\bibfnamefont {S.~V.}\ \bibnamefont
  {Streltsov}},\ }\href {\doibase 10.1021/acs.chemrev.0c00579} {\bibfield
  {journal} {\bibinfo  {journal} {Chemical Reviews}\ }\textbf {\bibinfo
  {volume} {121}},\ \bibinfo {pages} {2992} (\bibinfo {year}
  {2020})}\BibitemShut {NoStop}%
\bibitem{Peng2020} Y. Y. Peng, R. Fumagalli, Y. Ding, M. Minola, S. Caprara, D. Betto, M. Bluschke, G. M. De Luca, K. Kummer, E. Lefran\c cois, M. Salluzzo, H. Suzuki, M. Le Tacon, X. J. Zhou, N. B. Brookes, B. Keimer, L. Braicovich, M. Grilli, and G. Ghiringhelli, \href{http://dx.doi.org/10.1038/s41563-018-0108-3}{Nature Materials {\bf 17}, 697 (2018)}.
\bibitem [{\citenamefont {Tam}\ \emph {et~al.}(2022)\citenamefont {Tam},
  \citenamefont {Zhu}, \citenamefont {Ayres}, \citenamefont {Kummer},
  \citenamefont {Yakhou-Harris}, \citenamefont {Cooper}, \citenamefont
  {Carrington},\ and\ \citenamefont {Hayden}}]{TamNatCom2022}%
  \BibitemOpen
  \bibfield  {author} {\bibinfo {author} {\bibfnamefont {C.~C.}\ \bibnamefont
  {Tam}}, \bibinfo {author} {\bibfnamefont {M.}~\bibnamefont {Zhu}}, \bibinfo
  {author} {\bibfnamefont {J.}~\bibnamefont {Ayres}}, \bibinfo {author}
  {\bibfnamefont {K.}~\bibnamefont {Kummer}}, \bibinfo {author} {\bibfnamefont
  {F.}~\bibnamefont {Yakhou-Harris}}, \bibinfo {author} {\bibfnamefont {J.~R.}\
  \bibnamefont {Cooper}}, \bibinfo {author} {\bibfnamefont {A.}~\bibnamefont
  {Carrington}}, \ and\ \bibinfo {author} {\bibfnamefont {S.~M.}\ \bibnamefont
  {Hayden}},\ }\href {\doibase 10.1038/s41467-022-28124-y} {\bibfield
  {journal} {\bibinfo  {journal} {Nature Communications}\ }\textbf {\bibinfo
  {volume} {13}},\ \bibinfo {pages} {570} (\bibinfo {year} {2022})}\BibitemShut
  {NoStop}%
\bibitem [{\citenamefont {Li}\ \emph {et~al.}(2021)\citenamefont {Li},
  \citenamefont {Zhang}, \citenamefont {Yilmaz}, \citenamefont {Pai},
  \citenamefont {Marvinney}, \citenamefont {Said}, \citenamefont {Yin},
  \citenamefont {Gong}, \citenamefont {Tu}, \citenamefont {Vescovo},
  \citenamefont {Nelson}, \citenamefont {Moore}, \citenamefont {Murakami},
  \citenamefont {Lei}, \citenamefont {Lee}, \citenamefont {Lawrie},\ and\
  \citenamefont {Miao}}]{LiPRX2021}%
  \BibitemOpen
  \bibfield  {author} {\bibinfo {author} {\bibfnamefont {H.}~\bibnamefont
  {Li}}, \bibinfo {author} {\bibfnamefont {T.~T.}\ \bibnamefont {Zhang}},
  \bibinfo {author} {\bibfnamefont {T.}~\bibnamefont {Yilmaz}}, \bibinfo
  {author} {\bibfnamefont {Y.~Y.}\ \bibnamefont {Pai}}, \bibinfo {author}
  {\bibfnamefont {C.~E.}\ \bibnamefont {Marvinney}}, \bibinfo {author}
  {\bibfnamefont {A.}~\bibnamefont {Said}}, \bibinfo {author} {\bibfnamefont
  {Q.~W.}\ \bibnamefont {Yin}}, \bibinfo {author} {\bibfnamefont {C.~S.}\
  \bibnamefont {Gong}}, \bibinfo {author} {\bibfnamefont {Z.~J.}\ \bibnamefont
  {Tu}}, \bibinfo {author} {\bibfnamefont {E.}~\bibnamefont {Vescovo}},
  \bibinfo {author} {\bibfnamefont {C.~S.}\ \bibnamefont {Nelson}}, \bibinfo
  {author} {\bibfnamefont {R.~G.}\ \bibnamefont {Moore}}, \bibinfo {author}
  {\bibfnamefont {S.}~\bibnamefont {Murakami}}, \bibinfo {author}
  {\bibfnamefont {H.~C.}\ \bibnamefont {Lei}}, \bibinfo {author} {\bibfnamefont
  {H.~N.}\ \bibnamefont {Lee}}, \bibinfo {author} {\bibfnamefont {B.~J.}\
  \bibnamefont {Lawrie}}, \ and\ \bibinfo {author} {\bibfnamefont
  {H.}~\bibnamefont {Miao}},\ }\href {\doibase 10.1103/PhysRevX.11.031050}
  {\bibfield  {journal} {\bibinfo  {journal} {Physical Review X}\ }\textbf
  {\bibinfo {volume} {11}},\ \bibinfo {pages} {031050} (\bibinfo {year}
  {2021})}\BibitemShut {NoStop}%
\bibitem [{\citenamefont {Sheldrick}(2008)}]{Sheldrick2008}%
  \BibitemOpen
  \bibfield  {author} {\bibinfo {author} {\bibfnamefont {G.~M.}\ \bibnamefont
  {Sheldrick}},\ }\href {\doibase 10.1107/S0108767307043930} {\bibfield
  {journal} {\bibinfo  {journal} {Acta Crystallographica Section A}\ }\textbf
  {\bibinfo {volume} {64}},\ \bibinfo {pages} {112} (\bibinfo {year}
  {2008})}\BibitemShut {NoStop}%
\bibitem{vaklav2014}V. Pet\v{r}\'{i}\v{c}ek, M. Du\v{s}ek, and L. Palatinus, \href{https://doi.org/10.1515/zkri-2014-1737}{Zeitschrift f\"{u}r Kristallographie - Crystalline Materials {\bf 229}, 345 (2014)}.
\bibitem [{\citenamefont {Louie}\ \emph {et~al.}(1979)\citenamefont {Louie},
  \citenamefont {Ho},\ and\ \citenamefont {Cohen}}]{louie79}%
  \BibitemOpen
  \bibfield  {author} {\bibinfo {author} {\bibfnamefont {S.~G.}\ \bibnamefont
  {Louie}}, \bibinfo {author} {\bibfnamefont {K.-M.}\ \bibnamefont {Ho}}, \
  and\ \bibinfo {author} {\bibfnamefont {M.~L.}\ \bibnamefont {Cohen}},\ }\href
  {\doibase https://doi.org/10.1103/PhysRevB.19.1774} {\bibfield  {journal}
  {\bibinfo  {journal} {Phys. Rev. B}\ }\textbf {\bibinfo {volume} {19}},\
  \bibinfo {pages} {1774} (\bibinfo {year} {1979})}\BibitemShut {NoStop}%
\bibitem [{\citenamefont {Meyer}\ \emph {et~al.}(1997)\citenamefont {Meyer},
  \citenamefont {Els\"{a}sser},\ and\ \citenamefont {F\"{a}hnle}}]{meyer97}%
  \BibitemOpen
  \bibfield  {author} {\bibinfo {author} {\bibfnamefont {B.}~\bibnamefont
  {Meyer}}, \bibinfo {author} {\bibfnamefont {C.}~\bibnamefont {Els\"{a}sser}},
  \ and\ \bibinfo {author} {\bibfnamefont {M.}~\bibnamefont {F\"{a}hnle}},\
  }\href@noop {} {\enquote {\bibinfo {title} {Fortran90 program for mixed-basis
  pseudopotential calculations for crystals},}\ } (\bibinfo {year} {1997}),\
  \bibinfo {note} {\.Max-Planck-Institut f\"{u}r Metallforschung, Stuttgart
  (unpublished)}\BibitemShut {NoStop}%
\bibitem [{\citenamefont {Hamann}\ \emph {et~al.}(1979)\citenamefont {Hamann},
  \citenamefont {Schl\"uter},\ and\ \citenamefont {Chiang}}]{haman79}%
  \BibitemOpen
  \bibfield  {author} {\bibinfo {author} {\bibfnamefont {D.~R.}\ \bibnamefont
  {Hamann}}, \bibinfo {author} {\bibfnamefont {M.}~\bibnamefont {Schl\"uter}},
  \ and\ \bibinfo {author} {\bibfnamefont {C.}~\bibnamefont {Chiang}},\ }\href
  {\doibase https://doi.org/10.1103/PhysRevLett.43.1494} {\bibfield  {journal}
  {\bibinfo  {journal} {Phys. Rev. Lett.}\ }\textbf {\bibinfo {volume} {43}},\
  \bibinfo {pages} {1494} (\bibinfo {year} {1979})}\BibitemShut {NoStop}%
\bibitem [{\citenamefont {Bachelet}\ \emph {et~al.}(1982)\citenamefont
  {Bachelet}, \citenamefont {Hamann},\ and\ \citenamefont
  {Schl\"uter}}]{bache82}%
  \BibitemOpen
  \bibfield  {author} {\bibinfo {author} {\bibfnamefont {G.~B.}\ \bibnamefont
  {Bachelet}}, \bibinfo {author} {\bibfnamefont {D.~R.}\ \bibnamefont
  {Hamann}}, \ and\ \bibinfo {author} {\bibfnamefont {M.}~\bibnamefont
  {Schl\"uter}},\ }\href {\doibase https://doi.org/10.1103/PhysRevB.26.4199}
  {\bibfield  {journal} {\bibinfo  {journal} {Phys. Rev. B}\ }\textbf {\bibinfo
  {volume} {26}},\ \bibinfo {pages} {4199} (\bibinfo {year}
  {1982})}\BibitemShut {NoStop}%
\bibitem [{\citenamefont {Vanderbilt}(1985)}]{vande85}%
  \BibitemOpen
  \bibfield  {author} {\bibinfo {author} {\bibfnamefont {D.}~\bibnamefont
  {Vanderbilt}},\ }\href {\doibase https://doi.org/10.1103/PhysRevB.32.8412}
  {\bibfield  {journal} {\bibinfo  {journal} {Phys. Rev. B}\ }\textbf {\bibinfo
  {volume} {32}},\ \bibinfo {pages} {8412} (\bibinfo {year}
  {1985})}\BibitemShut {NoStop}%
\bibitem [{\citenamefont {Perdew}\ \emph {et~al.}(1996)\citenamefont {Perdew},
  \citenamefont {Burke},\ and\ \citenamefont {Ernzerhof}}]{perde96}%
  \BibitemOpen
  \bibfield  {author} {\bibinfo {author} {\bibfnamefont {J.~P.}\ \bibnamefont
  {Perdew}}, \bibinfo {author} {\bibfnamefont {K.}~\bibnamefont {Burke}}, \
  and\ \bibinfo {author} {\bibfnamefont {M.}~\bibnamefont {Ernzerhof}},\ }\href
  {\doibase https://doi.org/10.1103/PhysRevLett.77.3865} {\bibfield  {journal}
  {\bibinfo  {journal} {Phys. Rev. Lett.}\ }\textbf {\bibinfo {volume} {77}},\
  \bibinfo {pages} {3865} (\bibinfo {year} {1996})}\BibitemShut {NoStop}%
\bibitem [{\citenamefont {Heid}\ and\ \citenamefont {Bohnen}(1999)}]{heid99}%
  \BibitemOpen
  \bibfield  {author} {\bibinfo {author} {\bibfnamefont {R.}~\bibnamefont
  {Heid}}\ and\ \bibinfo {author} {\bibfnamefont {K.-P.}\ \bibnamefont
  {Bohnen}},\ }\href {\doibase https://doi.org/10.1103/PhysRevB.60.R3709}
  {\bibfield  {journal} {\bibinfo  {journal} {Phys. Rev. B}\ }\textbf {\bibinfo
  {volume} {60}},\ \bibinfo {pages} {R3709} (\bibinfo {year}
  {1999})}\BibitemShut {NoStop}%
\bibitem [{\citenamefont {Feng}\ \emph {et~al.}(2012)\citenamefont {Feng},
  \citenamefont {Wang}, \citenamefont {Jaramillo}, \citenamefont {van Wezel},
  \citenamefont {Haravifard}, \citenamefont {Srajer}, \citenamefont {Liu},
  \citenamefont {Xu}, \citenamefont {Littlewood},\ and\ \citenamefont
  {Rosenbaum}}]{Feng2012}%
  \BibitemOpen
  \bibfield  {author} {\bibinfo {author} {\bibfnamefont {Y.}~\bibnamefont
  {Feng}}, \bibinfo {author} {\bibfnamefont {J.}~\bibnamefont {Wang}}, \bibinfo
  {author} {\bibfnamefont {R.}~\bibnamefont {Jaramillo}}, \bibinfo {author}
  {\bibfnamefont {J.}~\bibnamefont {van Wezel}}, \bibinfo {author}
  {\bibfnamefont {S.}~\bibnamefont {Haravifard}}, \bibinfo {author}
  {\bibfnamefont {G.}~\bibnamefont {Srajer}}, \bibinfo {author} {\bibfnamefont
  {Y.}~\bibnamefont {Liu}}, \bibinfo {author} {\bibfnamefont {Z.-A.}\
  \bibnamefont {Xu}}, \bibinfo {author} {\bibfnamefont {P.~B.}\ \bibnamefont
  {Littlewood}}, \ and\ \bibinfo {author} {\bibfnamefont {T.~F.}\ \bibnamefont
  {Rosenbaum}},\ }\href {\doibase 10.1073/pnas.1202434109} {\bibfield
  {journal} {\bibinfo  {journal} {Proceedings of the National Academy of
  Sciences}\ }\textbf {\bibinfo {volume} {109}},\ \bibinfo {pages} {7224}
  (\bibinfo {year} {2012})}\BibitemShut {NoStop}%
\bibitem [{\citenamefont {Currat}(2006)}]{Currat2006}%
  \BibitemOpen
  \bibfield  {author} {\bibinfo {author} {\bibfnamefont {R.}~\bibnamefont
  {Currat}},\ }\enquote {\bibinfo {title} {Three-axis inelastic neutron
  scattering},}\ in\ \href {\doibase 10.1007/1-4020-3337-0_12} {\emph {\bibinfo
  {booktitle} {Neutron and X-ray Spectroscopy}}},\ \bibinfo {editor} {edited
  by\ \bibinfo {editor} {\bibfnamefont {F.}~\bibnamefont {Hippert}}, \bibinfo
  {editor} {\bibfnamefont {E.}~\bibnamefont {Geissler}}, \bibinfo {editor}
  {\bibfnamefont {J.~L.}\ \bibnamefont {Hodeau}}, \bibinfo {editor}
  {\bibfnamefont {E.}~\bibnamefont {Leli{\`e}vre-Berna}}, \ and\ \bibinfo
  {editor} {\bibfnamefont {J.-R.}\ \bibnamefont {Regnard}}}\ (\bibinfo
  {publisher} {Springer Netherlands},\ \bibinfo {address} {Dordrecht},\
  \bibinfo {year} {2006})\ pp.\ \bibinfo {pages} {383--425}\BibitemShut
  {NoStop}%
\end{thebibliography}

\end{document}